\documentclass[usenatbib]{mn2e}
\usepackage{graphicx}
\usepackage{epstopdf}
\usepackage{amssymb}
\usepackage{longtable}
\usepackage{dpfloat}
\usepackage{subfig}
\usepackage{journal_names}
\usepackage{xcolor} 

\title[Stream in NGC~4651]{Kinematics and simulations of the stellar stream in the halo of the Umbrella Galaxy}
\author[C. Foster et al.]{C. Foster,$^{1,2}$\thanks{E-mail: cfoster@aao.gov.au} H. Lux,$^{3,4}$ A.J. Romanowsky,$^{5,6}$ D. Mart\'inez-Delgado,$^{7}$ S. Zibetti,$^{8}$   \newauthor J.A. Arnold,$^9$ J.P. Brodie,$^{6,9}$  R. Ciardullo,$^{10}$ R.J. GaBany,$^{11}$  M.R. Merrifield,$^{3}$ 
\newauthor N. Singh,$^{12}$   and J. Strader$^{13}$
\\
$^1$Australian Astronomical Observatory, PO Box 915, North Ryde, NSW 1670, Australia\\
$^2$European Southern Observatory, Alonso de Cordova 3107, Vitacura, Santiago, Chile\\
$^3$School of Physics and Astronomy, The University of Nottingham, University Park, Nottingham, NG7 2RD, UK\\
$^4$Department of Physics, University of Oxford, Denys Wilkinson Building, Keble Road, Oxford, OX1 3RH, UK\\
$^5$Department of Physics and Astronomy, San Jos\'e State University, One Washington Square, San Jose, CA 95192, USA\\
$^6$University of California Observatories, 1156 High Street, Santa Cruz, CA 95064, USA\\
$^{7}$Astronomisches Rechen-Institut, Zentrum fur Astronomie der Universitat Heidelberg, Monchhofstr. 12-14, 69120, Heidelberg, Germany\\
$^8$INAF - Osservatorio Astrofisico di Arcetri, Largo Enrico Fermi 5, I-50125 Firenze - Italy\\
$^9$Department of Astronomy and Astrophysics, University of California, Santa Cruz, CA 95064, USA\\
$^{10}$Department of Astronomy \& Astrophysics, The Pennsylvania State University, 525 Davey Lab, University Park, PA 16802 USA\\
$^{11}$Black Bird Observatory, Mayhill, New Mexico, USA\\
$^{12}$Centre for Astronomy, National University of Ireland, Galway, University Road, Galway, Ireland\\
$^{13}$Department of Physics and Astronomy, Michigan State University, East Lansing, MI 48824, USA\\
}

\def\oiii{[O\,{\sc iii}]}
\def\hii{H\,{\sc ii}}
\def\hi{H\,{\sc i}}

\begin{document}

\date{}

\pagerange{\pageref{firstpage}--\pageref{lastpage}} \pubyear{2014}

\maketitle

\label{firstpage}

\begin{abstract}
We study the dynamics of faint stellar substructures around the Umbrella Galaxy, NGC 4651, which hosts a dramatic system of streams and shells formed through the tidal disruption of a nucleated dwarf elliptical galaxy. We elucidate the basic characteristics of the system (colours, luminosities, stellar masses) using multi-band Subaru/Suprime-Cam images. The implied stellar mass-ratio of the ongoing merger event is $\sim$\,1:50. We identify candidate kinematic tracers (globular clusters, planetary nebulae, \hii\ regions), and follow up a subset with Keck/DEIMOS spectroscopy to obtain velocities. We find 15 of the tracers are likely associated with halo substructures, including the probable stream progenitor nucleus. These objects delineate a kinematically cold feature in position--velocity phase space. We model the stream using single test-particle orbits, plus a rescaled pre-existing N-body simulation. We infer a very eccentric orbit with a period of $\sim$\,0.35 Gyr and turning points at $\sim$\,2--4 and $\sim$\,40 kpc, implying a recent passage of the satellite through the disc, which may have provoked the visible disturbances in the host galaxy. This work confirms that the kinematics of low surface brightness substructures can be recovered and modelled using discrete tracers -- a breakthrough that opens up a fresh avenue for unraveling the detailed physics of minor merging. 
\end{abstract}


\section{Introduction}

While classically pictured as detached ``island universes'', galaxies are now appreciated as actively connected to their environments, and growing continuously through the infall of smaller galaxies.  This view is supported both by discoveries of stellar substructures in their outer regions or ``haloes''  (e.g. \citealt{Searle78,Malin83}) and by the paradigm of hierarchical assembly in a cold dark matter (CDM) cosmology (e.g. \citealt{White78,Cooper11}). Consequently, the study of halo substructure has become a major industry (e.g. \citealt{Ibata01,Johnston08,Martell10,Helmi11,Mouhcine11,Xue11,Bate14}), both as a route to understanding the assembly histories of galaxies, and as a proving ground for CDM via quantitative tests of substructure predictions.

These tests to date have turned up a number of problems that are variations of the longstanding small-scale substructure ``crisis''  (\citealt{Moore94,Klypin99}; see review in \citealt{Weinberg14}). These include deficits of satellites around the Milky Way and other galaxies, and shallow dark matter cores rather than central cusps in galaxies over a wide range of luminosities (e.g. \citealt{deBlok08,Herrmann09,Agnello12,Newman13b}). These puzzles might be resolved through better modelling of baryon physics, or by recourse to CDM alternatives -- but distinguishing between these two general options could be an insurmountable obstacle for analyses that focus only on the central regions of galaxies.

A different, and arguably more robust, line of inquiry is to study the large-scale galaxy properties that are minimally affected by baryon physics and so may be more uniquely compared to theoretical predictions. These properties include the orbits and total (virial) masses of satellites. Recent work along these lines include comparisons of the orbital coherence of satellite groups to predictions for
cosmological infall (e.g. \citealt{Pawlowski12,Wang13}), and searches for the impact signatures of small dark matter subhaloes on visible stellar streams \citep{Carlberg13}.

The investigations of substructure have so far been dominated by studies of the two main galaxies in the Local Group (M31 and the Milky Way), but  it is critically important to extend them to more distant galaxies -- both to build up better statistics, and to understand how the properties of substructure depend systematically on galaxy type and environment. Although progress has been made along these lines (see \citealt{Atkinson13} and references therein), it has been almost exclusively based on photometry, while missing the dimension of velocity that is obtained directly through spectroscopy. Such information is indispensable for inferring the three-dimensional (3D) anatomy of substructures, the timescales of orbital decay, the progenitor properties including total mass, and the dynamical interplay between satellite and host galaxies. While these aspects can sometimes be probed through gas dynamics (e.g. \citealt{Iodice03}), most infalling satellites are gas poor, in a reflection of the morphology--density relation that is probably caused by quenching during the accretion process (e.g. \citealt{Einasto74, Mayer06, Grcevich10, Geha12, Slater13, SanchezJanssen13}).

The paucity of gas in halo substructures motivates pursuing stellar-light spectroscopy, where the challenge is the extremely low surface brightnesses of typically $\sim$~27 mag arcsec$^{-2}$ or fainter. An alternative in such cases is to exploit bright, discrete tracer objects: globular clusters (GCs) and planetary nebulae (PNe). This approach is now used extensively for chemodynamical mapping of the faint outer regions of early-type galaxies (e.g. \citealt{Coccato09,Usher12,Pota13a}), but so far has had very limited application to visible substructures.  Pioneering examples include PN- and GC-based spectroscopic studies of tidal debris around M51 \citep{Durrell03}, of an outer halo stream in M87 \citep{Romanowsky12}, of substructures around M31 \citep{Veljanoski13a}, and of a tidal stream linking NGC~4365 and NGC~4342 \citep{Blom14}. Other substructures have been found serendipitously in large kinematic datasets \citep{Merrett03,McNeil10,Shih10,Ventimiglia11,Cortesi11}, but such cases are hard to model without photometric counterparts for reference.

The time is ripe for a systematic spectroscopic survey of distant halo substructures, if it can be demonstrated that the observations are feasible, and that the subsequent dynamical modelling and interpretation is tractable. To this end, here we present a pilot study of the `Umbrella Galaxy', NGC~4651, using multi wavelength photometry, spectroscopy of GCs, PNe, and \hii\ regions, and preliminary dynamical modelling of its substructure.

\begin{figure*}
\begin{center}
\includegraphics[width=150mm]{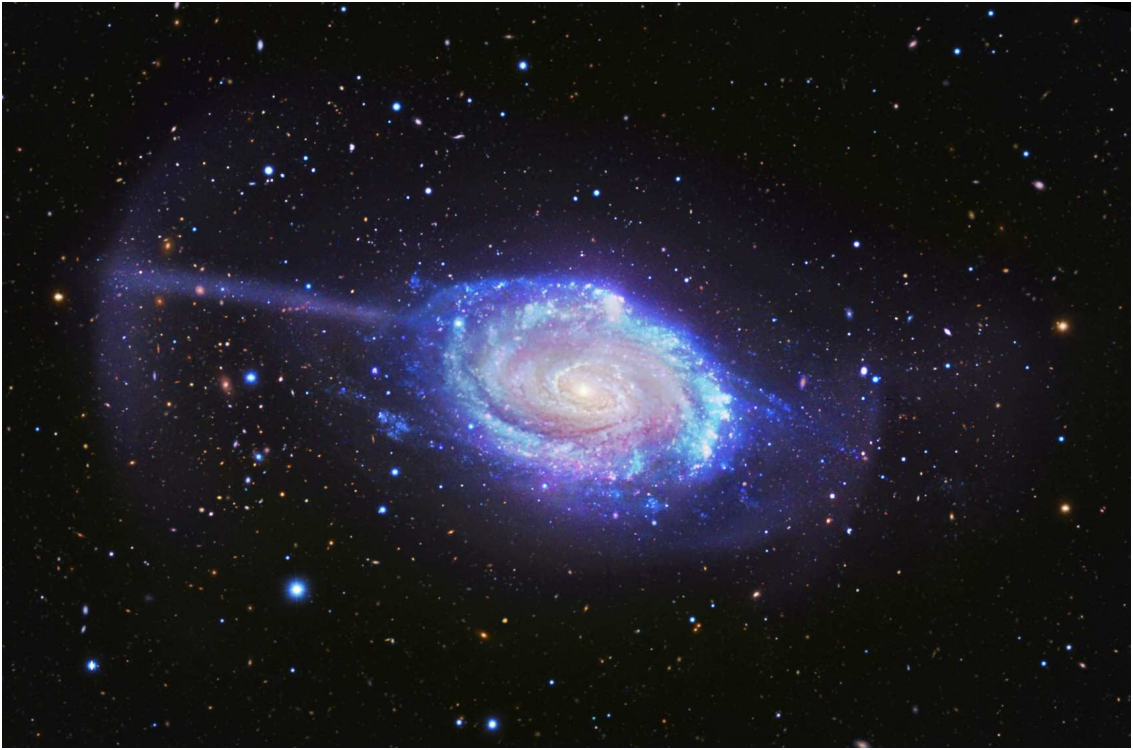}
\caption{Overview of the Umbrella Galaxy NGC~4651 with its surrounding stellar substructures. This colour composite image was produced by combining luminance-filter data from the 0.5~m BlackBird Remote Observatory telescope with $g'$, $r'$, and \oiii\ images from Suprime-Cam on the 8.2~m Subaru telescope in green, red and blue, respectively. A narrow stream protrudes to the left and terminates at a perpendicular arc.
On the right are two shell-like features, and beyond them an even fainter, irregular protrusion. Most of the features are consistent with multiple wrappings of a single tidal stream, except for the protrusion at far right which likely consists of kicked-out disc stars. North is up and East to the left.
}\label{fig:prettypic}
\end{center}
\end{figure*}

NGC~4651 is a spiral galaxy (Sc) that hosts one of the brightest and most spectacular systems of halo substructures in the local Universe, as seen in Fig.~\ref{fig:prettypic}. A dramatic umbrella-like feature extends far beyond the disc to the East side of the galaxy, and consists of a straight, collimated stream (henceforth, stick) terminating in a sharp, perpendicular plume (henceforth, shell). On the West side there are several broad plume- and shell-like features, whose morphologies are more difficult to distinguish owing to their lower surface brightnesses and overlapping locations.

The umbrella feature was first reported by \citet[where it was interpreted as a dwarf companion]{Zwicky56}, and was discussed in various subsequent catalogues including \citet{Arp66}. The main galaxy disc appears relatively undisturbed, with a normal star formation rate \citep{Kennicutt98}. The outer disc shows mild kinematic disturbances in ionized gas \citep{Rubin99} and a very lopsided  H~{\sc i} distribution, with a tail-like feature to the West that is coincident with a broad optical plume, and {\it no} detectable gas to the East in the umbrella region \citep{Chung09}.

In a prelude to the current paper, \citet{MartinezDelgado10} acquired deep optical imaging of NGC~4651 as part of a survey of eight nearby Milky Way-like spiral galaxies with suspected halo substructures, using a small robotic telescope. They described the morphology of the umbrella and counter-shell, and noted similarities to simulations of stellar halo accretion events from \citet{Johnston08}. They inferred that the observed features could plausibly have formed in the tidal disruption of a low-mass galaxy on a near radial orbit around 6 to 10 Gyr ago.

The initial goals of our follow-up project are to quantify the luminosity, colour, and stellar mass of the substructure; to measure its kinematical properties; to derive an orbital model that accounts for the essential features of the observations; and to interpret the results in the context of cosmological accretion predictions.

\begin{figure*}
\begin{center}
a) Cleaned $i'$-band image\\\includegraphics[width=150mm]{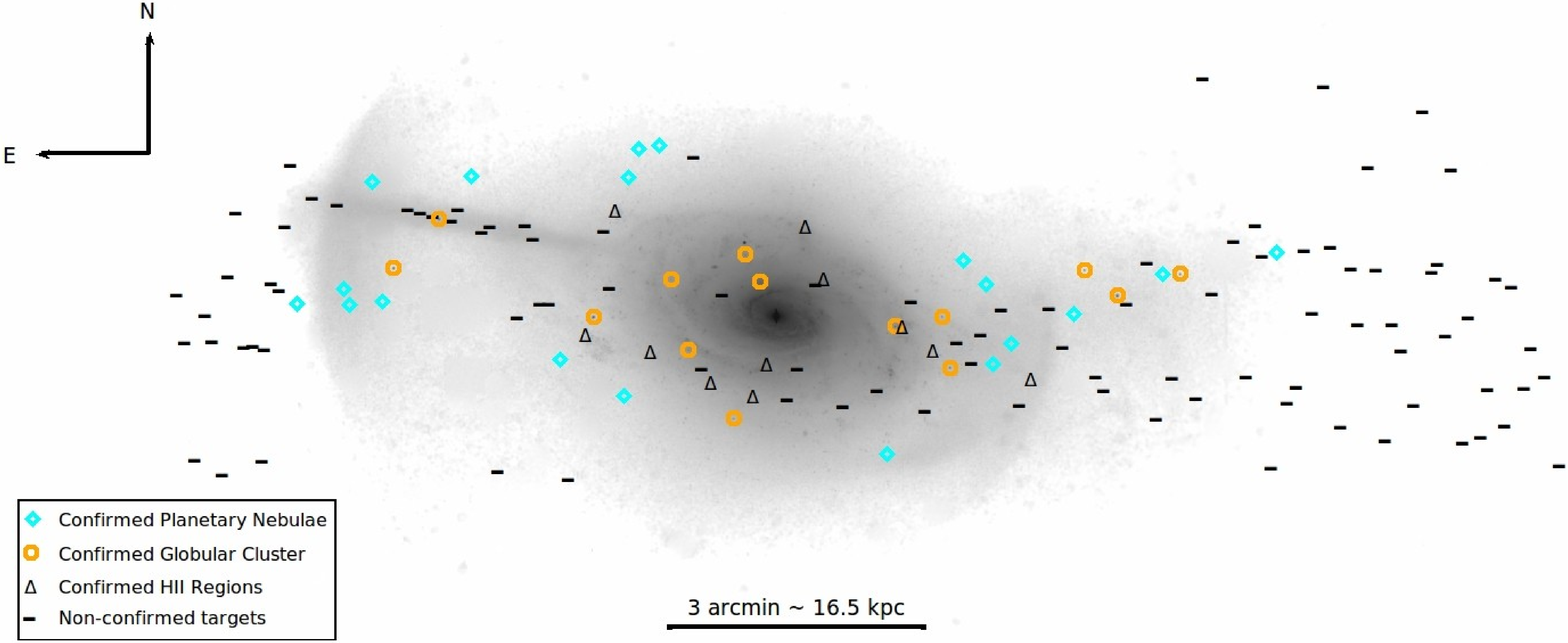}\\
b) Stellar mass map\\\includegraphics[width=150mm]{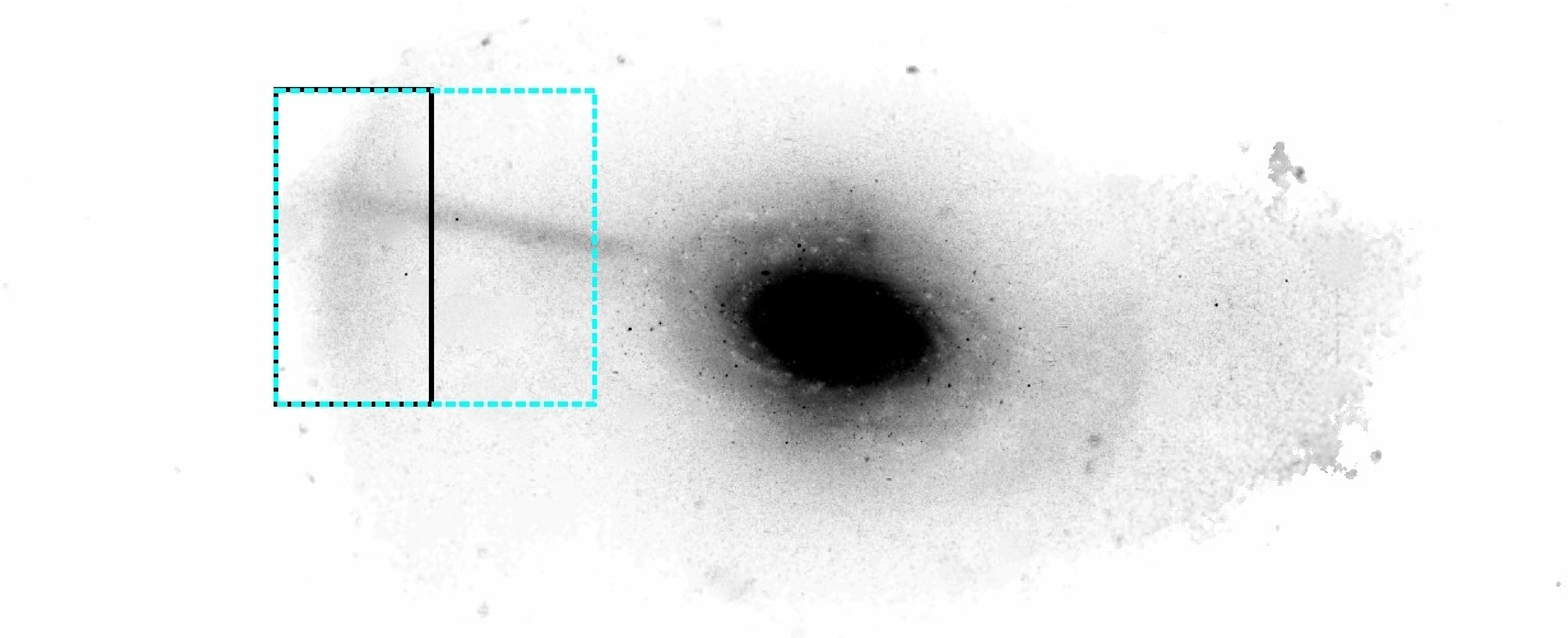}\\
c) $NUV$ image\\\includegraphics[width=150mm]{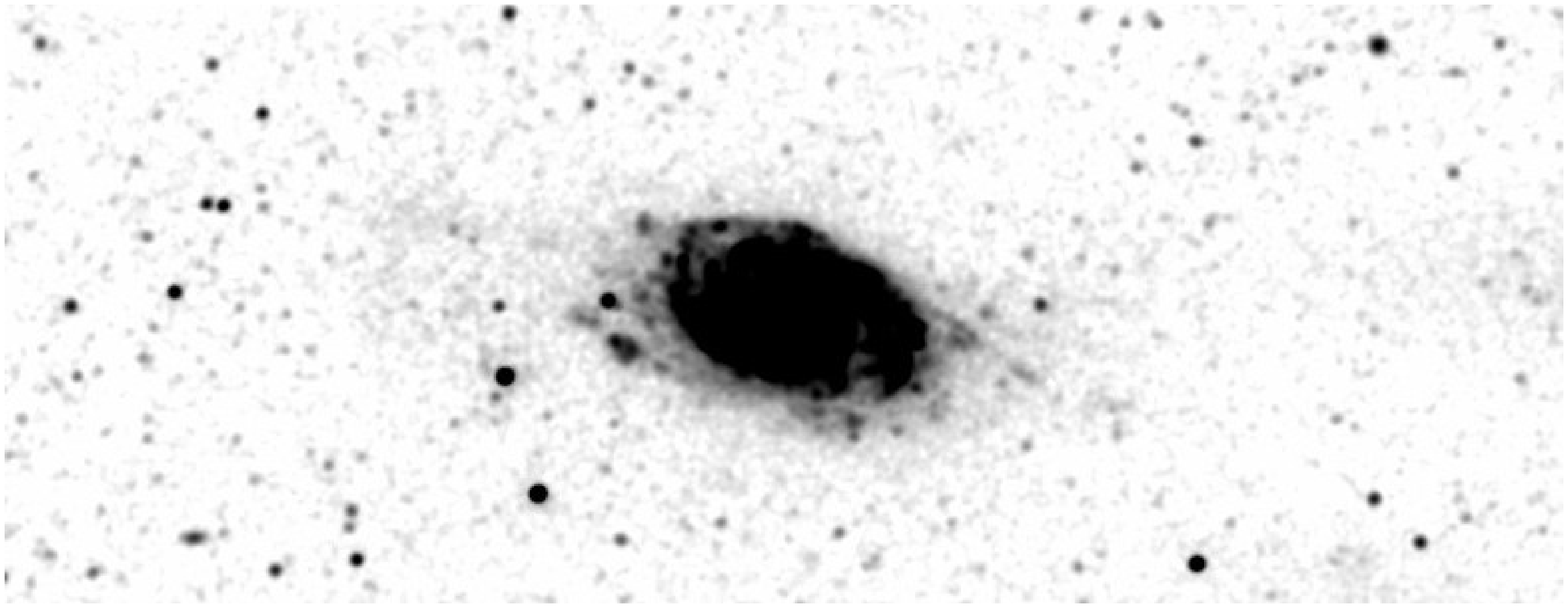}
\caption{Substructures around NGC~4651, with North up and East to the left as shown. A scale is provided for reference. The upper panel (a) is the $i'$-band image with contamination subtracted and levels adjusted to highlight substructures. Surface brightness levels range from $\mu_i \sim$~27 to $\sim$~19~mag~arcsec$^{-2}$.  Orange circles and black triangles show the positions of confirmed GCs and \hii\  regions, while cyan diamonds show confirmed member PNe. Horizontal dashes show the positions of other spectroscopically observed candidates. Panel (b) is the inferred stellar mass map, with solid black and dashed cyan rectangles highlighting regions used for estimating the stellar mass in the umbrella (see Section \ref{sec:images}). The stellar surface density levels displayed range from $\Sigma_\star \sim 3\times10^5$ to $\sim 3\times10^7 M_\odot$~kpc$^{-2}$. The lower panel (c) shows the archival {\it GALEX} $NUV$ image, smoothed with a 3 pixel Gaussian (4.5 arcsec) on a matching spatial scale. }\label{fig:targets}
\end{center}
\end{figure*}

Various properties of NGC~4651 are summarized in Table~\ref{tab:prop}. Of special note is the distance, which
was not well-established in the literature, and which we estimate as 19 Mpc
based on the planetary nebula luminosity function (see Section \ref{sec:dist}). The implied angular scale is 92 pc arcsec$^{-1}$.

\begin{figure*}
\begin{center}
\includegraphics[width=.95\textwidth]{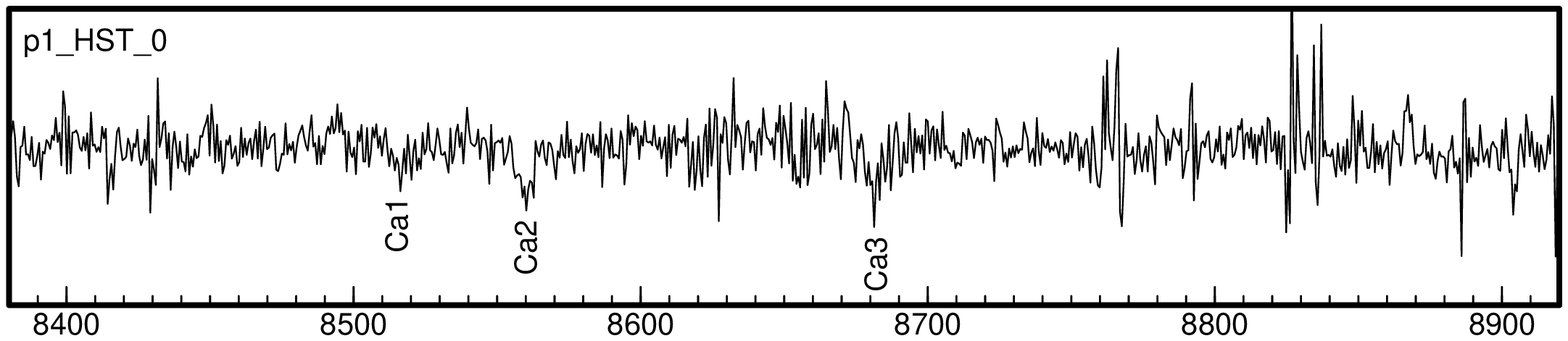}\\\includegraphics[width=0.95\textwidth]{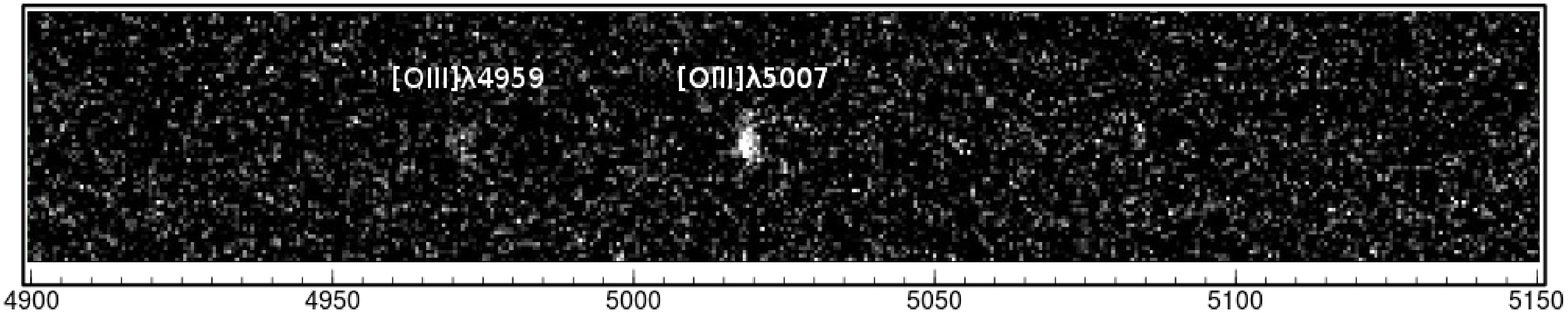}\\\includegraphics[width=0.95\textwidth]{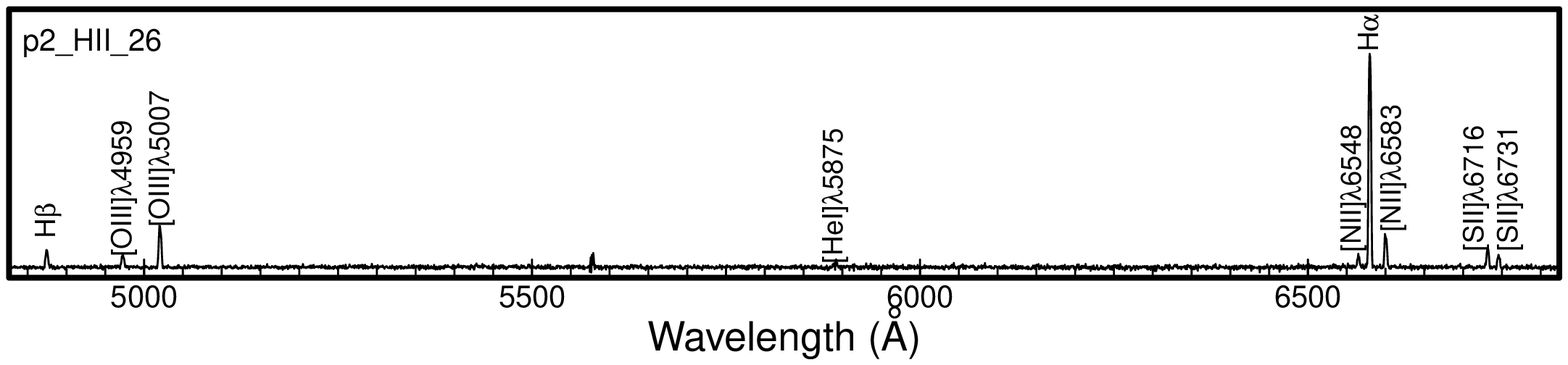}
\caption{Examples of typical DEIMOS spectra, for a GC  (1D spectrum; upper panel),  a PN (2D spectrum; middle panel) and an \hii\ region (1D spectrum; lower panel). Salient features are labelled. The PN spectrum (p1\_PN\_14) is centred on the clear \oiii\ $\lambda$5007 emission line, with wavelength and spatial axes along the $x$ and $y$ axes, respectively. As several other PN spectra, a hint of \oiii\ $\lambda$4959 detection is visible.}\label{fig:spec}
\end{center}
\end{figure*}

\begin{table*}
\begin{center}
\caption{Values (column 2) of salient measurable properties of NGC~4651\ and its substructure from various sources (column 3).}\label{tab:prop}
\begin{tabular}{lll}
\hline\hline
Property&Value&Reference\\
(1)&(2)&(3)\\
\hline
{\it NGC~4651}\\
Distance &19 Mpc&This work\\
Total extinction-corrected $B$-band apparent magnitude &$11.04$ mag&RC3, \citealt{deVaucouleurs91}\\
Inclination& $53\pm2$ degrees&\citealt{Epinat08}\\
Deprojected rotation amplitude&$215\pm10$  km s$^{-1}$&\citealt{Epinat08}\\
$B$-band 25th magnitude isophote diameter &244.4 arcsec&RC3, \citealt{deVaucouleurs91}\\
Heliocentric recession velocity &$795\pm5$ km s$^{-1}$&This work\\
$H$-band bulge-to-total luminosity ratio & 0.36 & \citealt{Laurikainen04} \\
$H$-band disc scale length & 26.4 arcsec & \citealt{Laurikainen04} \\
$H$-band bulge effective radius & 17.2 arcsec & \citealt{Laurikainen04} \\
$i'$-band absolute magnitude / luminosity & $-21.3 \pm 0.1$ mag / $(2.1 \pm 0.2)\times10^{10} L_\odot$ & This work\\
Colour $(g'-i')_0$ & $0.80 \pm 0.15$ mag& This work\\
Stellar mass & $(1.7\pm0.7) \times10^{10} M_\odot$ & This work\\
\hi\ gas mass & $(5.7\pm0.3)\times10^{9} M_\odot$ & distance-scaled from \citealt{Chung09} \\
\hline
{\it Substructure}\\
Umbrella $i'$-band absolute magnitude / luminosity & $-16.7 \pm 0.1$ mag / $(3.2\pm0.3)\times10^8 L_\odot$ & This work\\
Umbrella colour $(g'-i')_0$ & $0.57 \pm 0.15$ mag & This work\\
Umbrella stellar mass &  $(1.8\pm0.7)\times10^8 M_\odot$& This work\\
\hline
\end{tabular}
\end{center}
\end{table*}

This paper is organized as follows: in Section \ref{sec:data}, we present the new imaging and spectroscopic data. Section \ref{sec:analysis} presents our analysis, while dynamical modelling is undertaken in Section \ref{sec:models}. Some implications are discussed in Section~\ref{sec:discussion}, and our conclusions can be found in Section \ref{sec:conclusions}.

\section{Data}\label{sec:data}

In an effort to better define the properties of the faint substructures, new images were obtained on the nights of 2011 January 3 and  4 in photometric conditions using the Suprime-Cam wide-field camera ($34 \times 27\,{\rm arcmin}^2$ field-of-view) on the 8.2 m Subaru telescope \citep{Miyazaki02}. NGC 4651 was imaged under 0.9, 0.6 and 0.9 arcsec  seeing in the $g'$ (1300 sec), $r'$ (500 sec), and $i'$-band (505 sec) filters, respectively, using a preset dithering pattern that offset successive images in order to compensate for gaps between Suprime-Cam's 10 CCDs (laid out in a 5$\times$2 grid).  An \oiii\ narrow band image (4860 sec)  was also acquired for  \hii\ region and PN candidate selection, under 1.2 arcsec seeing. The images were reduced using a customized version of the SDFRED2 pipeline (\citealt{Ouchi04}; bias subtraction, flat-fielding, and distortion correction) and astrometric solutions were obtained using SDSS point-source catalogs. The individual frames in each filter were then projected to a common astrometric coordinate system and co-added to create final mosaics.

As is common practice \citep[e.g.][]{Foster11}, candidate GCs were selected based on their positions in colour-colour space using $g'$, $r'$ and $i'$-band photometry. The GC candidates are dominated by a peak of blue objects at $(g'-i') \sim 0.8$, with a secondary peak at $(g'-i') \sim 1.1$. The surface density profile of candidates reaches the background level at a galoctocentric radius of $\sim$~6 arcmin ($\sim$~30~kpc), inside of which there are $\sim$~140~GCs down to $i=24$. Table \ref{table:GC} reports basic photometric data of all GC candidates.

\begin{table*}
\begin{center}
\caption{Catalog of photometrically selected globular cluster candidates (column 1) in order of priority. Columns 2 and 3 give the position in right ascencion and declination (J2000), respectively. Columns 4 to 11 are the Subaru/Suprime-Cam photometry. Spectroscopically confirmed globular clusters are marked with an asterix (*). This table is available in full online as supplementary material.}
\begin{tabular}{lllrlrlrlrl}
\hline
Name & $\alpha$ & $\delta$ & $g$ & $g_{\rm err}$ & $r$ & $r_{\rm err}$ & $$ & $i_{\rm err}$ & $m(5007)$ & $m(5007)_{\rm err}$\\
(1)&(2)&(3)&(4)&(5)&(6)&(7)&(8)&(9)&(10)&(11)\\
\hline
  p1\_HST\_0* & 12:43:36.783 & 16:23:28.827 & 21.51 & 0.01 & 21.003 & 0.0060 & 20.692 & 0.0060 & 23.946 & 0.103\\
  p1\_HST\_1 & 12:43:41.611 & 16:22:58.854 & 23.151 & 0.186 & 22.467 & 0.077 & 22.164 & 0.069 & 25.085 & 0.108\\
  p1\_HST\_2 & 12:43:39.348 & 16:22:32.539 & 23.092 & 0.038 & 22.287 & 0.016 & 21.852 & 0.012 & 25.671 & 0.114\\
  p1\_HST\_3 & 12:43:40.733 & 16:23:58.736 & 20.203 & 0.02 & 20.042 & 0.016 & 20.771 & 0.035 & 21.094 & 0.1\\
  p1\_sup\_0 & 12:43:17.302 & 16:12:20.677 & 21.947 & 0.0020 & 21.583 & 0.0030 & 21.471 & 0.0020 & 24.497 & 0.105\\
  ... & ... & ... & ... & ... & ... & ... & ... & ... & ... & ...\\
 \hline
\end{tabular}
\label{table:GC}
\end{center}
\end{table*}	

\begin{table*}
\small
\begin{center}
\caption{Catalog of photometrically selected candidate \hii\, regions (column 1) ordered by priority. Columns 2 and 3 give the position in right ascencion and declination (J2000), respectively. Columns 4 to 11 are the Subaru/Suprime-Cam photometry. Spectroscopically confirmed \hii regions are marked with an asterix (*). This table is available in full online as supplementary material.}
\begin{tabular}{lllrlrlrlrl}
\hline
Name & $\alpha$ & $\delta$ & $g$ & $g_{\rm err}$ & $r$ & $r_{\rm err}$ & $$ & $i_{\rm err}$ & $m(5007)$ & $m(5007)_{\rm err}$\\
(1)&(2)&(3)&(4)&(5)&(6)&(7)&(8)&(9)&(10)&(11)\\
\hline
  p2\_HII\_0 & 12:43:58.439 & 16:24:40.433 & 24.418 & 0.062 & 24.117 & 0.039 & 24.019 & 0.042 & 25.037 & 0.108\\
  p2\_HII\_1 & 12:43:47.364 & 16:24:36.712 & 23.315 & 0.03 & 23.122 & 0.016 & 24.164 & 0.055 & 24.401 & 0.104\\
  p2\_HII\_2 & 12:43:46.291 & 16:23:53.180 & 27.559 & 7.806 & 25.738 & 1.65 & 25.508 & 0.817 & 25.44 & 0.111\\
  p2\_HII\_3 & 12:43:37.564 & 16:24:00.320 & 23.511 & 0.115 & 23.576 & 0.094 & 23.876 & 0.131 & 24.57 & 0.105\\
  p2\_HII\_4 & 12:43:48.174 & 16:24:29.292 & 21.993 & 0.013 & 21.702 & 0.0090 & 22.689 & 0.021 & 22.955 & 0.101\\
  ... & ... & ... & ... & ... & ... & ... & ... & ... & ... & ...\\
\hline
\end{tabular}
\label{table:HII}
\end{center}
\end{table*}	

\begin{table}
\begin{center}
\caption{Catalog of photometrically selected planetary nebula candidates (column 1) in order of priority. Columns 2 and 3 give the position in right ascencion and declination (J2000), respectively. Columns 4 and 5 are the Subaru/Suprime-Cam \oiii\, photometry. Spectroscopically confirmed planetary nebulae are marked with an asterix (*). This table is available in full online as supplementary material.}
\begin{tabular}{lllrlrlrlrl}
\hline
Name & $\alpha$ & $\delta$ & $m(5007)$ & $m(5007)_{\rm err}$\\
(1)&(2)&(3)&(4)&(5)\\
\hline
  p1\_PN\_0 & 12:43:36.042 & 16:11:02.532 & 27.242 & 0.159\\
  p1\_PN\_1 & 12:43:38.714 & 16:35:54.191 & 28.43 & 0.277\\
  p1\_PN\_2 & 12:44:10.423 & 16:23:18.293 & 27.858 & 0.205\\
  p1\_PN\_3 & 12:43:18.252 & 16:21:48.844 & 27.814 & 0.2\\
  p1\_PN\_4 & 12:44:06.596 & 16:25:23.775 & 27.744 & 0.194\\
 ...& ... & ... & ... & ...\\
\hline
\end{tabular}
\label{table:PNe}
\end{center}
\end{table}	

The PNe and \hii\ regions were first identified by eye based on their \oiii\ $\lambda$5007 emission on the narrow-band image. The selection was refined based on their \oiii\ -to-optical colours (cf.\ \citealt{Arnaboldi02}). Candidate \hii\ regions and PNe were differentiated based on the lack of detectable continuum flux in the broadband images in the latter. Tables \ref{table:HII} and \ref{table:PNe} report basic photometric data for \hii\, region and PN candidates.

Follow-up spectra of GCs, PNe and \hii\ regions were simultaneously obtained on the night of 2011 March 31 using a multi-slit mask with the DEep Imaging Multi-object Spectrograph (DEIMOS; \citealt{Faber03}) on the Keck telescope under 0.9--1.0 arcsec seeing conditions. Extra calibrations (blue arcs and flats) were obtained on the night of 2011 December 2. The GG455 filter and 600 l mm$^{-1}$ grating centred on 6900 \AA\space combination was used together with a 1 arcsec slit width. This setup yields a nominal resolution of $\Delta\lambda \sim 3.5$ \AA\space and allows for a generous wavelength coverage ($\sim$4300--9600 \AA). Two 15-minute exposures were taken, yielding a modest total time on target of 0.5 hours. Fig. \ref{fig:targets} (top) shows the positions of the science target slits. Out of the 37 PN candidates in the immediate vicinity of NGC~4651, slits were placed on 27 of them, and velocities were successfully measured for 19 objects. Most of the GCs with recovered velocities have magnitudes of $i \sim$~21--22.

In addition, a longslit spectrum along the stream (see Section \ref{sec:kin}) was obtained with Keck/DEIMOS on 12 Apr 2013. The 1200 l mm$^{-1}$ grating was used, centred at 7800 \AA\, with the 1 arcsec slit. This setup gives a resolution of 1.5 \AA. Three 20-min exposures were taken, with the goal of measuring the velocities both of the putative stream nucleus, and of the stream itself (the former was successful).

All DEIMOS data were reduced using the {\sc idl spec2d} data reduction pipeline provided online \citep{Cooper12, Newman13a}. Flat-fielding was performed using internal flats. Separate
flats were used for the blue and red chips for the multiple-slits observation. The wavelength calibration for the multiple-slits data  was carried out using HgCdKrArZn and ArKrNeXe arc lamps for the blue and red chips, respectively, while long-slit wavelength calibration used ArKrNeXe arcs only. Local sky subtraction was performed within the pipeline. Fig. \ref{fig:spec} shows typical sections of spectra for a GC, a PN and an \hii\ region.

\section{Analysis and results}\label{sec:analysis}

Here we present our main observational results, beginning in Section~\ref{sec:dist} with a measurement of the planetary nebulae luminosity function (PNLF) to obtain a more precise distance for NGC~4651. We  describe our methods for obtaining the kinematics of the various tracers in \ref{sec:tracerskin}, and perform a photometric analysis of the stream in Section~\ref{sec:images}. We use a kinematic analysis to assign tracers to disc and stream populations in Section~\ref{sec:kin}. We summarise the set of stream tracers and specific frequency considerations in Section~\ref{sec:streamtracers}.

\subsection{Distance}\label{sec:dist}

Existing estimates for the distance to NGC~4651 show considerable variation, ranging between 12 and 16~Mpc based on the redshift, and $15$ to $33\,{\rm Mpc}$ from the Tully--Fisher relation \citep{Willick97,Ekholm00}. Fortunately, our new imaging data allow us to greatly improve the distance estimate to a precision of $\sim$~10\% by analysis of the PNLF.

We begin by calibrating our narrow-band \oiii\ observations using images of the standard star GD~248, and converting the resulting AB magnitudes to monochromatic $\lambda 5007$ flux using the filter transmission curve of \citet{Arnaboldi03}, an assumed heliocentric radial velocity of $788$~km~s$^{-1}$ \citep{Epinat08}, and the procedures outlined in \citet{Jacoby87} and \citet{Jacoby89b}.  Following \citet{Jacoby89a}, Table~\ref{table:values} lists these fluxes in terms of magnitudes where
\begin{equation}
m_{5007} = -2.5 \log F_{5007} - 13.74
\end{equation}
with $F_{5007}$ in units of ergs~cm$^{-2}$~s$^{-1}$.

\begin{figure}
\begin{center}
\includegraphics[width=\columnwidth]{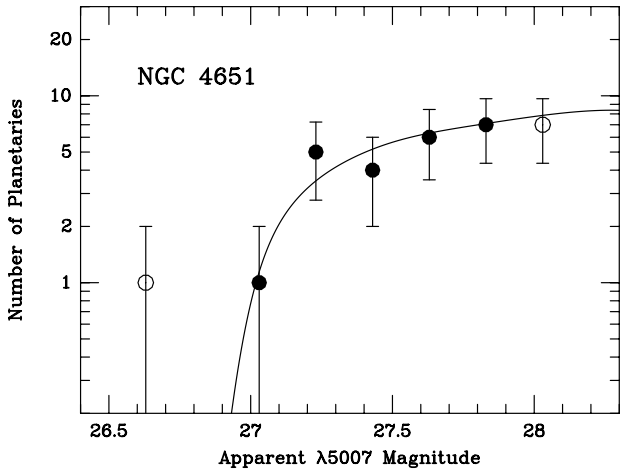}
\caption{The planetary nebula luminosity function around NGC~4651, binned into 0.2~mag intervals.  Open circles represent PNe that are either ``overluminous'' or beyond the completeness limit.  The curve shows the best-fit empirical PNLF convolved with the photometric error function.  The best-fit distance modulus is $(m-M)_0 = 31.36^{+0.06}_{-0.11}$.}\label{fig:pnlf}
\end{center}
\end{figure}

Fig.~\ref{fig:pnlf} shows the resulting PNLF, using all 31 photometric PN candidates in the vicinity of NGC~4651 with good photometry (as will be discussed in Section~\ref{sec:pnkin}, the vast majority of these appear to be valid PNe from spectroscopy; one likely background galaxy, PN\_6, has been omitted from the PNLF fitting). Apart from PN\_9, which is more than 0.3~mag more luminous than any other source, the shape of the luminosity function above $m_{5007} = 27.9$ agrees extremely well with the empirical law proposed by \citet{Ciardullo89}:
\begin{equation}
N(M) \propto {\rm e}^{0.307 M} \left\{ 1 - {\rm e}^{3 (M^* - M)} \right\}.
\label{eq:pnlf}
\end{equation}
A K--S test shows that the distribution of all PN candidates (except for PN\_9) is consistent with being drawn from the universal PNLF. This indicates that contamination by e.g. \hii\ regions is unlikely to be significant. \citet{Ciardullo02} has shown that for systems more metal-rich than the Large Magellanic Cloud, the bright-magnitude cut-off is $M^* = -4.47$, regardless of the age of the parent stellar population. We can therefore fit the observed PNLF to this function via the method of maximum likelihood \citep{Ciardullo89}, and obtain an independent estimate of the distance to the system.

The curve in Fig.~\ref{fig:pnlf} shows the best-fit PNLF of equation~\ref{eq:pnlf} convolved with a series of Gaussian kernels representing the effect of the photometric errors.  For a foreground extinction of $E(B-V) = 0.023$, the distance modulus of NGC~4651 is $(m-M)_0 = 31.36^{+0.06}_{-0.11}$, or $18.7^{+0.6}_{-1.0}$~Mpc, where the uncertainties represent only the formal statistical error of the fit. To this value one should add in quadrature an additional error of $\sim 0.07$~mag, which represents the uncertainties in the photometric zero point, our knowledge of the filter transmission function, and the error in the foreground reddening.  Note that if we restrict the PNLF to only those \oiii\ $\lambda 5007$ sources spectroscopically confirmed as PNe, the distance to the system does not change significantly, with $D = 18.8^{+0.7}_{-1.2}$~Mpc.

Our PNLF distances comes with some caveats.  The first, of course, involves the fact that the fit shown in Fig.~\ref{fig:pnlf} does not include the brightest \oiii\ source, PN\_9. ``Overluminous'' \oiii\ sources are not uncommon in PN surveys beyond $\sim 15$~Mpc (for example, 11 have been identified in the Virgo Cluster; \citealt{Jacoby90}), through their presence is still somewhat of a mystery.  Fortunately, these objects stand out from the bulk of the PN population:  if one were to include PN\_9 in the statistical sample, the overall fit to the empirical PNLF would be much worse (excluded by a K--S test), and the resulting distance would be $\sim 0.3$~mag closer.

The second caveat concerns the metallicity of the system.  We will see in Section~\ref{sec:images} that the PNe belonging to the stream derive from an underlying stellar population with $g \sim 14.6$ (i.e. twice the luminosity of the umbrella).  If we use the $g'$-band to $V$-band transformation of \citet{Jester05} and apply a bolometric correction of $-0.85$ \citep{Buzzoni06}, then the maximum-likelihood fit described above yields a luminosity-specific PN density $\alpha_{0.5} = 14.3^{+3.4}_{-2.5} \times 10^{-9}$~PN~$L_{\odot}^{-1}$ for objects within 0.5~mag of $M^*$.  This number is much higher than that observed for the old, metal-rich stellar populations of elliptical galaxies, and is larger than the $\alpha$ values associated with spiral galaxy discs \citep{Ciardullo05, Ciardullo10}. It is, however, marginally consistent with the high PN densities seen in the small Local Group galaxies such as M32, NGC~185 and NGC~205 (\citealt{Corradi05}; see Section~\ref{sec:streamtracers} for further discussion). It is also expected that some of the PN are associated with NGC~4651 itself, thereby artificially enhancing $\alpha$. Indeed, in Section \ref{sec:streamtracers} we show that 47\% of the PNe have kinematics consistent with that of the disc of NGC~4651.

The association of the PN with a small, disrupted galaxy is consistent with our dynamical analysis (see below).  However, it also suggests that our PNLF distance may be overestimated:  as demonstrated by \citet{Ciardullo92} and \citet{Ciardullo02}, $M^*$ fades in systems with metallicities below [O/H] $< 8.5$.  It is therefore possible that our distance is overestimated by $\sim 0.1$~mag due to this effect.

Finally, there is the ongoing uncertainty about the zero point of the PNLF method.  Many of the galaxies targeted by the PNLF also have distances from the Surface Brightness Fluctuation (SBF) method \citep{Tonry01, Blakeslee10b}, but a comparison of the two datasets has shown a $\Delta(m-M)_0 \sim 0.35$~mag offset between the two methods.  The most plausible explanation for this offset is the existence of a very small amount of internal reddening ($E(B-V) \sim 0.02$) in the $\sim 6$ spiral bulges which serve to calibrate the two techniques \citep{Ciardullo02}.  If this reddening is greater than that found in normal elliptical galaxies, both methods would be affected, but in opposite directions:  SBF distances would be systematically overestimated, while PNLF distances would be systematically underestimated.  Hence, it is possible that the PNLF distances to elliptical galaxies and other non-dusty systems need to be increased by $\sim 0.2$~mag.

The GCs provide two other constraints on the distance, in principle. First is the GC luminosity function (GCLF), whose turnover magnitude may be used as a standard candle to first order \citep[see e.g.][and references therein]{Rejkuba12}. A detailed analysis of the GCLF, including careful completeness corrections, is beyond the scope of this paper, but a preliminary check suggests a distance in the range of $\sim$~20--30~Mpc.

Second is the GC apparent sizes, since the intrinsic sizes of $\sim$~2--4~pc also provide a standard ruler (e.g. \citealt{Larsen03,Alexander13}). We have measured half-light radii for four confirmed GCs in archival $F555W$ \emph{Hubble Space Telescope}/WFPC2 images using \emph{ishape} \citep{Larsen99}. Additional details on the fitting process can be found in \citet{Strader11}. We find mean half-light radii of 0.037 arcsec with random uncertainties around 10--15\%. This implies a distance of 14--17 Mpc assuming a mean intrinsic size of 2.5--3 pc.

In more detail, we can use the apparent colours of the GCs to correct their apparent magnitudes for extinction from the galaxy disc (a relatively small effect), and consider where they would reside in a size--luminosity plane \citep{Brodie11} for a series of different distances. The plausible range of distances thus obtained is $\sim$~12--20~Mpc, with larger distances implying a peculiar population of large and bright objects like $\omega$~Cen.

These two rough GC-based analyses suggest a distance of $\sim$~20~Mpc, which is consistent with the PNLF result above, and we therefore adopt a distance of 19~Mpc throughout this paper. In general, the qualitative results are not sensitive to this assumption, and derived masses and distances can be scaled for an alternative distance.

\subsection{Kinematics}\label{sec:tracerskin}

Because we are using a variety of kinematic tracers, recession velocities for each type are measured using different methods depending on what is most appropriate. Below, we give a description of the respective methods. Once the recession velocity of a given target is secured, we separate targets associated with NGC~4651 from background galaxies and foreground stars based on their recession velocities. We consider any object part of the NGC~4651 system if it has $600 \le V_{\rm obs} \le 1100$ km s$^{-1}$. The nearest contaminants are very well separated in velocity space with a gap of $\gtrsim350$ km s$^{-1}$ from the nearest confirmed target.

\subsubsection{Globular clusters}

The GC kinematics are measured using {\sc fxcor} in {\sc iraf}. We perform a cross-correlation of the spectra around the Calcium Triplet spectral features (centred around rest wavelengths of 8498, 8542 and 8662 \AA) with a sample of six template spectra. We use template spectra from the single stellar population models of \citet{Vazdekis10} after adjusting the resolution to match the science spectra using Gaussian convolution. We choose models of varying metallicities ranging from $\rm[Fe/H]=-1.7$ to 0.22 dex and a single age of 10 Gyr, i.e. typical stellar population parameters for GCs \citep[e.g.][]{Puzia05,Strader05,Norris08}. We also confirm less certain objects with $\rm H\alpha \, \lambda6563$. The values quoted for the observed recession velocity ($V_{\rm obs}$) of GCs given in Table \ref{table:values} are the average from the 6 templates. The uncertainties represent the average of the output errors given by {\sc fxcor}, added in quadrature to the standard deviation of $V_{\rm obs}$ between templates.

We use the higher-resolution long-slit spectrum to confirm the radial velocity of the putative stream nucleus. We cross-correlate the spectrum with a set of templates in the region around the Calcium Triplet (which is similar to the procedure used in \citealt{Pota13a} for GCs). We find a heliocentric radial velocity for this object of $712\pm14$ km s$^{-1}$, which is consistent with the lower-resolution result of $742\pm22$ km~s$^{-1}$, and we use the latter in the rest of this paper for self-consistency of the entire kinematic data set. The long-slit spectrum also yielded a velocity for the bright blue object near the point where the stream crosses the disc, and we determine that this object is a foreground star.

\subsubsection{\hii\ region kinematics}\label{sec:HIIkin} 

We measure the recession velocity of \hii\ regions using the {\sc iraf} procedure {\sc rvidlines}. We provide a list of the visible strong emission lines in the spectrum, then manually identify a few lines, and the procedure finds the others. A total of between 7 and 9 emission lines are used for our \hii\ region spectra. The quoted recession velocity and uncertainty (see Table \ref{table:values}) are the output from {\sc rvidlines} based on the positions of identified emission lines.

\subsubsection{PN kinematics}\label{sec:pnkin}

We confirm PNe and measure their velocities based on a clear signal in the \oiii\ $\lambda$5007 emission line (in some cases, the weaker \oiii\ $\lambda$4959 line is also detected). Out of the 27 PN candidates observed, 19 have clear \oiii\ detections, with most of the remaining objects showing some marginal evidence for detection such that they might well be recovered with longer exposure times. This high success rate demonstrates the robustness of the PN selection procedures, and the relative ease of spectroscopically recovering emission line objects once they have been detected by narrow-band imaging.

We use the {\sc iraf} routine {\sc imexam} to interactively fit the \oiii\ line on the 2D reduced spectra. The measured wavelength of the line is then simply converted to a recession velocity. We use the relationship between the peak counts in the \oiii\ line as a function of recession velocity uncertainty, as measured for the \hii\ regions (see Section \ref{sec:HIIkin}), to estimate uncertainties for the PNe.

As a sanity check, we co-add all 1-D spectra of PN candidates and measure the ratio of \oiii\ $\lambda$5007 to $\rm H\alpha \, \lambda6563$ to be $\approx2.9$. We find that ratio to be consistent with expectations for PNe. This is a further indication that contamination by e.g. \hii\ regions is not significant.

\begin{figure}
\begin{center}
\includegraphics[width=93mm]{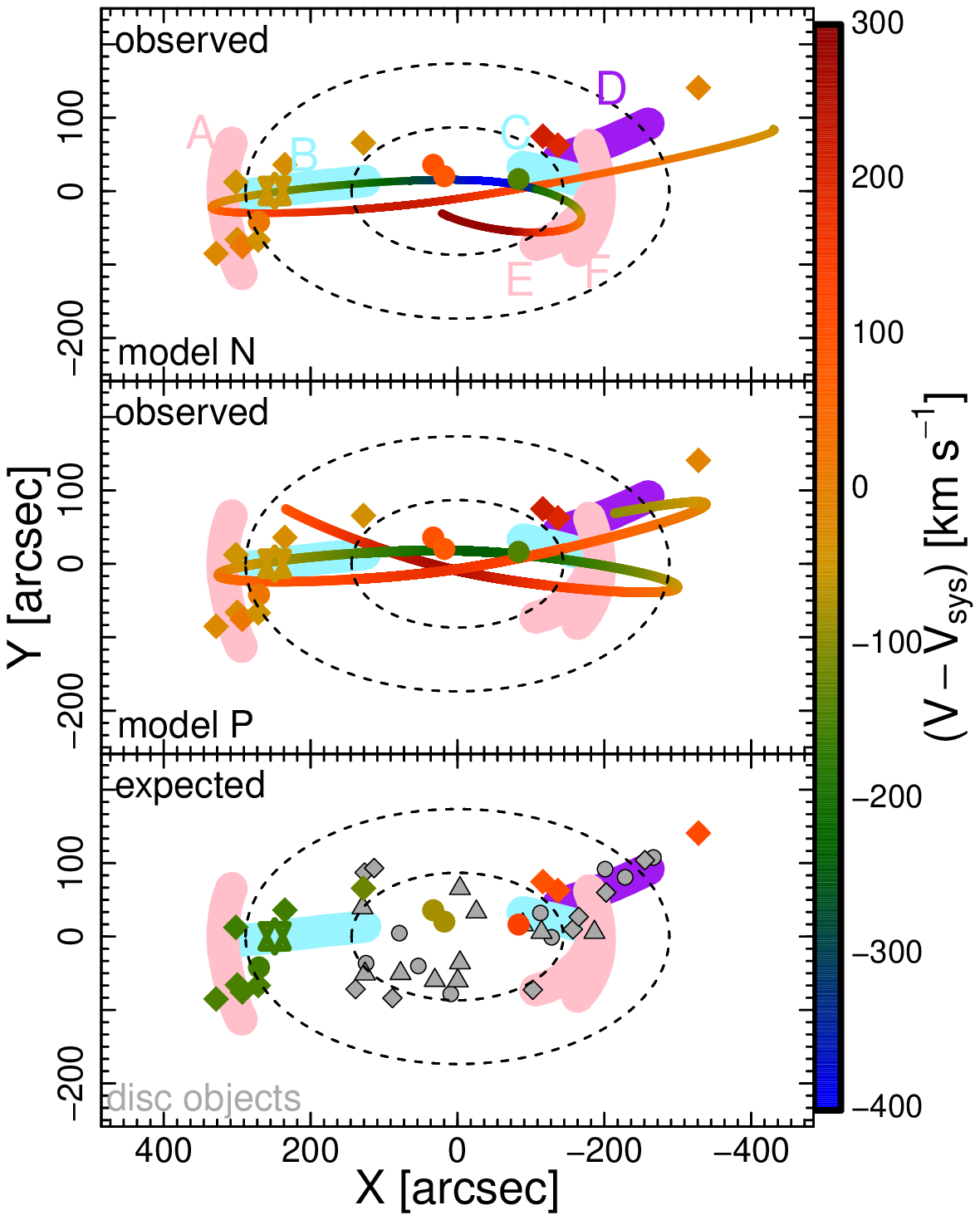}
\caption{Spatial distribution and kinematics of GCs (circles), PNe (diamonds) and \hii\ regions (triangles), using the observed (upper and middle panels) or expected (lower panel) velocities based on a disc model. X and Y run parallel to the major and minor axes of NGC~4651, respectively, and are measured relative to $\alpha=$12:43:42.65 and $\delta=$16:23:36.0. Symbol colours correspond to relative velocity ($V-V_{\rm sys}$), with grey symbols used for disc tracers in the lower panel only. The tentatively identified nucleus of the infalling galaxy is shown with a six-pointed star. The cyan regions of the stream (B,C) were matched, the pink shells (A, E, F) were not, and the purple plume (D) was used to limit the range of model parameters. Good match models for the parameter ranges of sets N and P as described in Table \ref{table:MCMCparams} are shown in the upper and middle panels, respectively. These involve the nucleus residing on the near- and far-side of the galaxy in the line-of-sight direction, respectively.  The two models provide a fair representation of the stream morphology and of the non-disc tracer kinematics. We cannot exclude that there are other models that match the data as well or better.}\label{fig:spatial}
\end{center}
\end{figure}

\begin{table*}
\begin{center}
\caption{Total foreground-dust corrected $g'$, $r'$ and $i'$ band magnitudes, $(g'-i')$ colours and stellar mass of NGC~4651 (column 2), the stick (column 3), shell (column 4) and umbrella (i.e., stick + shell, column 5). Uncertainties may well be  underestimated due to the complications associated with background estimates and the very rough definitions of the various components. See text for detail.}\label{tab:phot}
\begin{tabular}{ccccc}
\hline\hline
 &	NGC~4651&		Stick&	Shell&	Umbrella\\
 &(2) &(3)&(4)&(5)\\
\hline
$g'_0$ [mag]&10.9$\pm$0.1&16.3$\pm$0.1&15.9$\pm$0.1&15.3$\pm$0.1\\
$r'_0$ [mag]&10.3$\pm$0.1&15.8$\pm$0.1&15.5$\pm$0.1&14.9$\pm$0.1\\
$i'_0$ [mag]&10.1$\pm$0.1&15.6$\pm$0.1&15.3$\pm$0.1&14.7$\pm$0.1\\
$(g'-i')_0$ [mag]&$0.80\pm$0.15&0.70$\pm$0.15&0.69$\pm$0.15&0.69$\pm$0.15\\
$M/L_i$ [$M/L_{\odot,i}$] & $0.79\pm0.28$ & $0.57\pm0.20$ & $0.57\pm0.20$ & $0.57\pm0.20$ \\
Stellar mass [$M_\odot$] & $(1.7\pm0.7)\times10^{10}$ & $(0.8\pm0.2)\times10^8$ & $(1.1\pm0.4)\times10^8$ & $(1.8\pm0.7)\times10^8$ \\
\hline
\end{tabular}
\end{center}
\end{table*}

\vspace{1cm}

\noindent In summary, we have spectroscopically confirmed 14, 19, 11 of the 56, 27 and 17 GC, PN and \hii\ region candidates, respectively, as belonging to the potential well of NGC~4651. We have also found 6 foreground stars with recession velocities $<230$ km s$^{-1}$ and 3 potential background galaxies.

\begin{figure*}
\begin{center}
\includegraphics[width=0.8\textwidth]{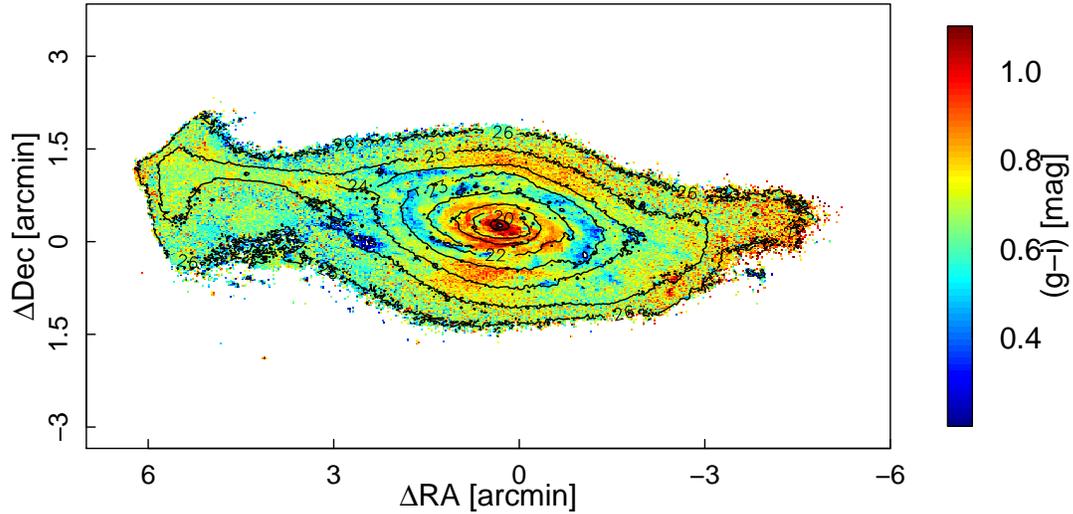}
\caption{Observed, dereddened $(g'-i')$-colour map of NGC~4651 for $\mu_r<26$ mag, based on the Suprime-Cam images, with overlaid $r'$-band isophotes (labelled black thin lines). The colour of the substructures is intermediate (yellow-green, $g'-i' \sim 0.7$) between that of the central bulge (orange-red, $g'-i' \sim$~0.9--1.0) and the spiral arms (cyan-blue, $g'-i' \sim$~0.4--0.5). The red colour on the far right is an artifact of residuals from the image reduction. Axes are computed from $\alpha=$12:43:42.65 and $\delta=$16:23:36.0.}\label{fig:gi}
\end{center}
\end{figure*}

\subsection{Photometry of the stream}\label{sec:images} 

In order to obtain ``clean'' maps of surface brightness, colours and stellar mass, we process the Subaru images as follows.  Removal of foreground Galactic stars and background objects such as GCs and galaxies is achieved in a two-step process. First, we perform ``unsharp masking'' filtering on the combined images in the three bands ($g,r,i$), to highlight the presence of point-like sources and small galaxies: a median-smoothed version of the image is subtracted from the full resolution one. SExtractor \citep{Bertin96} is then run to detect all significant sources and output a source map. From this map, a mask is created, which is broadened to cover the extent of the point spread function wings to our surface brightness limit. \textsc{fixpix} in IRAF is used to interpolate over the masked pixels in each of the images. We exclude the central, high surface brightness regions of the galaxy from the process, because this would result in interpolating real structure in NGC\,4651. As a second step, bright foreground stars with extended haloes of scattered light are manually edited by replacing those regions with local background and noise, using the \textsc{imedit} task in IRAF. The same is done for bright foreground stars in the central regions.

Background subtraction is performed in each band by fitting a second degree 2D polynomial in a rectangular annulus approximately 8 arcmin around the galaxy. Residual large scale background fluctuations are estimated as the RMS of the median background level in a large number of boxes approximately 1--2 arcmin wide, randomly distributed around the galaxy. These levels correspond to 30.2, 29.5, and 29.2 mag arcsec$^{-2}$ in $g'$, $r'$ and $i'$ band respectively (corrected for Galactic extinction using \citealt{Schlegel98}).

In order to ensure reliable colour measurements pixel by pixel, we use \textsc{adaptsmooth} \citep{Zibetti09a,Zibetti09b} to make an adaptive median smoothing of the images such that a minimum signal-to-noise ratio of 20 is reached at any pixel in all three bands simultaneously. Only those pixels that are at least 10-$\sigma$ of the background fluctuation levels in all bands are retained. Finally, we convolve the $r'$- and $i'$-band images with a Gaussian kernel to match the poorest seeing of 0.93 arcsec in the $g'$-band image.

Based on the maps obtained in this way, we compute local colours and local stellar mass-to-light ratio $M_\star/L_i$ from $(g'-i')$, following the prescriptions given in Appendix~B of \cite{Zibetti09a}, where a \citet{Chabrier03} stellar initial mass function was adopted. Note that although there is a classic degeneracy in inferring age and metallicity from a single colour, the inferred $M_\star/L$ value is more robust. From these data we derive the stellar mass map shown in the middle panel of Fig.~\ref{fig:targets}. 

Within the cyan box, we integrate the stellar mass with surface brightnesses $\mu_r\le24.5$, which corresponds to the umbrella stick, and estimate that it contains $(0.8\pm0.2)\times10^8 M_\odot$ with $(M/L)_g=0.64$ and $(M/L)_i=0.57$. Similarly, we integrate the light within the black box corresponding to surface brightnesses of $24.5\le\mu_r\le25.5$ (i.e. the shell only), and obtain $(1.1\pm0.4)\times10^8 M_\odot$ with $(M/L)_g=0.63$ and $(M/L)_i=0.57$. Adding the two, the total mass of the umbrella is $(1.8\pm0.7)\times10^8 M_\odot$. Given the counter-shells seen on the opposite side of the galaxy, we estimate that the total stream mass is roughly double that of the umbrella, i.e. $\sim4\times10^8 M_\odot$.

In order to estimate the total mass of the main galaxy NGC~4651, we integrate the stellar mass corresponding to a surface brightness brighter than $\mu_r=25.5$ mag arcsec$^{-2}$, yielding $(1.7\pm0.7)\times10^{10}M_\odot$ with $(M/L)_g=0.96$ and $(M/L)_i=0.79$ (which is consistent with more simplified estimates based on overall $(B-V)$ colour in \citealt{TorresFlores11}). Hence the merger that produced the substructure has a ratio of $\sim$~1:30 to 1:40 in $i'$-band luminosity, and 
$\sim$~1:50 in stellar mass ($\mu_*\sim 0.02$). Magnitudes for the various components of the stream and NGC~4651 can be found in Table \ref{tab:phot}. 

The umbrella stick is $\sim$~4~arcmin ($\sim$~20~kpc) long, although it may have a longer extent that crosses the face of the galaxy disc. Its full-width at half-maximum is $\simeq 2.2\pm0.6$~arcsec, or $\simeq 200\pm 60$~pc. The umbrella shell spans $\sim$~3.5~arcmin ($\sim$~20~kpc), as do the shells on the West side.

\begin{figure*}
\begin{center}
\includegraphics[width=0.9\textwidth]{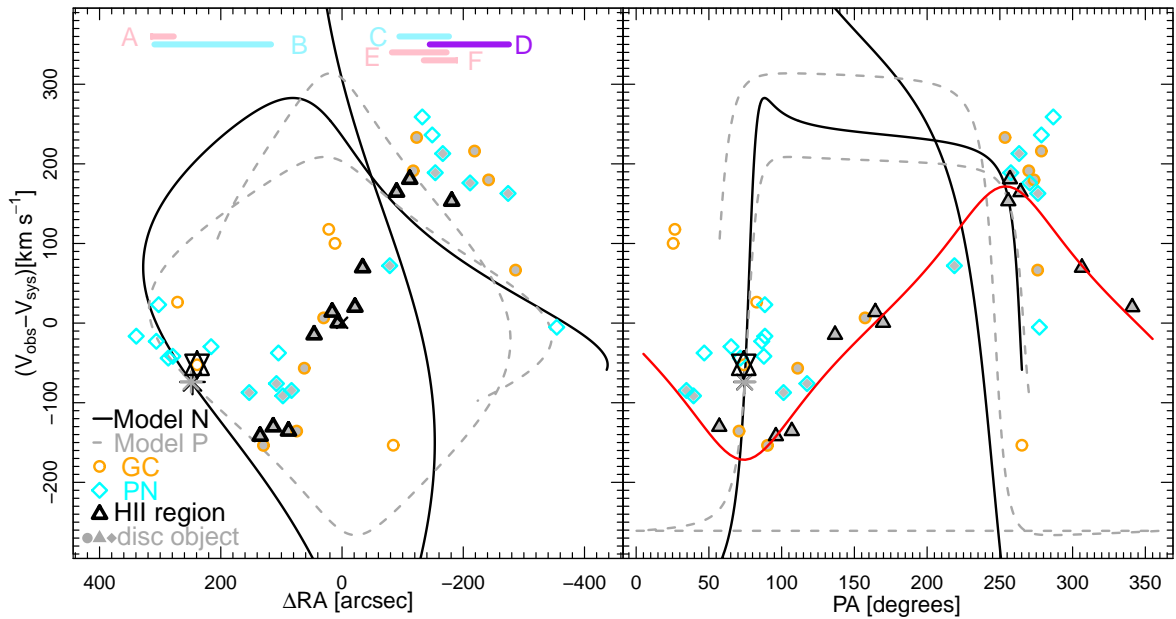}
\caption{Observed recession velocities with systemic velocity subtracted ($V_{\rm obs}-V_{\rm sys}$) of confirmed GCs (orange circles), \hii\ regions (black triangles) and PNe (cyan diamonds) as a function of differential right ascension from the galaxy centre ($\Delta RA$ from $\alpha=$12:43:42.65, left panel) and position angle (PA measured counterclockwise from North, right panel). Grey filled symbols have kinematics consistent with the disc of NGC~4651. The nucleus of NGC~4651 is shown as a black cross (left panel only) and  the possible infalling galaxy nucleus as a six-pointed star. The red solid curve shows the rotation of the disc of NGC~4651 as measured by  \citet{Epinat08}. Solid black and dashed grey curves show the single particle models N and P described in Section \ref{sec:models}, respectively, with their progenitor/nucleus position marked by an asterisk. The spatial ranges of substructures A, E, F (pink), B, C (cyan) and D (purple) defined in Fig.~\ref{fig:spatial} are in the upper left panel.
}\label{fig:rv}
\end{center}
\end{figure*}

In order to search for signs of recent star formation traced by H$\alpha$ emission, we use the composite stellar population synthesis libraries of Zibetti et al. (2009) to compute the
expected $(g'-r')$ colour for a given $(g'-i')$. A colour excess $\Delta(g'-r')=(g'-r')_{\rm observed} - (g'-r')_{\rm predicted}$ is computed at every pixel. After correcting for a small systematic offset between observations and models (derived for regions clean of H$\rm\alpha$ emission), we find no significant excess. 

The umbrella feature is extremely faint in the FUV, consistent with an old stellar population with no excess found in archival {\it Galaxy Evolution Explorer} ({\it GALEX}) imaging. However, there appears to be mild NUV excess to the South-West in a narrow arm-like feature that deviates from the main spiral arm angles and parallels our inferred stream trajectory (see Fig.~\ref{fig:targets} and cyan arm in Fig.~\ref{fig:spatial}). It is possible that this feature consists of a wake of star-forming regions provoked by the passage of the stream.
We do not detect any other sign of ongoing star formation in the substructure regions.

The colours of halo substructures can be used to connect them to their progenitor galaxy types (cf.\ \citealt{Rudick10,Ludwig12,Mihos13,Gu13}). Fig.~\ref{fig:gi} shows the $(g'-i')$ colour map from Suprime-Cam (we have also generated a shallower map with SDSS DR8, to  $\mu_{r'}\sim$~25 mag arcsec$^{-2}$, and found very good consistency). In this image, it is apparent that the colour of the umbrella and other substructures is intermediate between the redder inner disc/``bulge" and the bluer spiral arms, and similar to the inter-arm regions and to the outer disc. This intermediate colour is expected for a progenitor satellite that is both more metal-poor than its host galaxy, and older than the spiral arms (given its lack of gas and star-forming regions). Unfortunately the colour information is not helpful for determining the origin of the Western plume, owing to the very low surface brightness in that region, and to the similar colours of the stream and outer disc. It would also be useful to measure a colour difference between the umbrella and the surrounding diffuse stellar halo of NGC~4651 (cf.\ \citealt{Ibata14}), but our photometry does not go deep enough to make a clear estimate of the halo colour.

In summary, the photometric properties of the stream are consistent with a dwarf galaxy progenitor with an absolute magnitude of $M_V \sim -17.0$, which would correspond to an intermediate-luminosity dwarf elliptical. The typical half-light radius of such a dwarf would be $\sim$~1~kpc \citep{Norris14};
the much smaller width of the stream may be an effect of the stripping process.
The typical metallicity would be [Fe/H]~$\sim -0.9$~dex \citep{Woo08,Kirby13}, which in combination with the observed colour would imply an age of $\sim$~4~Gyr.

We note that an older, more metal-rich population as in the Sgr stream and the M31 Giant Southern Stream would have $(g'-i' )\sim$ 1.0--1.1, which is clearly excluded by our photometry. In fact, the blue colour of the stream (equivalent to $B-V \sim 0.65$) appears to be very unusual for a luminous dwarf elliptical, and is reminiscent of dwarfs with recent star formation, which are more common in low-density environments \citep{Gavazzi10,Kim10}, including NGC~205 in the Local Group \citep{Mateo98}.
Thus the Umbrella Galaxy system may provide an example of dwarf quenching during infall to a larger galaxy (see also \citealt{Beaton14}).

\subsection{Kinematic analysis}\label{sec:kin}

In order to distinguish objects associated with NGC~4651 from those that may belong to the infalling system, we compare their observed kinematics to those of the disc of NGC~4651 as measured by the GHASP survey in \citet{Epinat08}. The expected recession velocity ($V_{\rm  exp}$) of the $i^{\rm th}$ tracer located at a position angle $PA_i$ is described by the following equation for a tilted disc:
\begin{equation}\label{eq:Vobs2}
V_{{\rm exp},i}=V_{\rm sys} \pm \frac{V_{\rm rot}}{\sqrt{1+\left(\frac{\tan(PA_{i}-PA_{\rm kin})}{\cos(incl)}\right)^2}},
\end{equation}
where the ambivalent sign is positive if $(PA_{i}-PA_{\rm kin})$ is in the first or fourth quadrants, and negative if it lies in the second or third quadrants. The kinematic position angle $PA_{\rm 
kin}=74^\circ$, the inclination $incl=53^\circ$, and the reprojected rotational velocity $V_{\rm rot}=172$ km s$^{-1}$ are based on \citet{Epinat08}. For the systemic velocity ($V_{\rm sys}$), we use our measured value of 795 km s$^{-1}$ (see Table \ref{table:values} for NGC4651\_nuc). 

\begin{figure}
\begin{center}
\includegraphics[width=85mm]{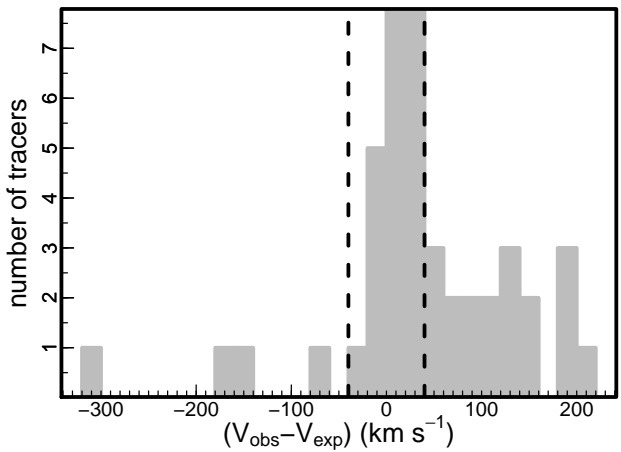}
\caption{Histogram of the difference between the observed and expected kinematics from the measured disc rotation at the position of each object. Objects which have $|V_{\rm obs}-V_{\rm exp}|<40$ km s$^{-1}$ (dashed lines) within the observational uncertainties are tagged as belonging to the disc of NGC~4651.}\label{fig:hist}
\end{center}
\end{figure}

The resulting disc rotation curve for a given value of $PA$ is shown in Fig.~\ref{fig:rv} (right panel). Fig.~\ref{fig:hist} shows a histogram of the difference between the expected line-of-sight velocity and that observed for all the identified kinematic tracers. Based on this distribution, we classify as ``disc'' objects those that have velocities consistent, within the uncertainties, with lying within 
$40\,{\rm km}\,{\rm s}^{-1}$ of the model disc velocity field at their location. 
These objects are highlighted in Fig.~\ref{fig:rv}.  We assume that all other objects do not belong to the disc, but to either the stream or a more general halo population.

Fig.~\ref{fig:spatial} shows the expected and observed 2D kinematics of these tracers.  Many tracers that coincide spatially with the obvious substructures seen in the Subaru image (labelled here with letters A--F) also emerge as kinematic outliers, supporting their association with the stream.
One of these objects, a ``GC'', 
is located directly on the umbrella stick (B). Its intermediate colour ($g'-i'=0.74$) and luminosity ($m_i=20.9$ mag or $M_i=-10.5$ mag) are consistent with that observed for dwarf elliptical nuclei in Virgo \citep[see][]{Brodie11}. For these reasons, we tentatively identify this object (p1\_sup\_79) as the possible nucleus of the shredded infalling system. 
We note that it is generally difficult to identify stream progenitors even with high quality images \citep[e.g.][]{MartinezDelgado12}. The likelihood of this object being the progenitor's nucleus is discussed in further detail in Section \ref{sec:models}.

The tracers in the umbrella also delineate a characteristic chevron-shaped feature in position--velocity phase-space which is the classic signature of a dynamically-cold shell on a radial orbit. We derive a rough internal velocity dispersion by computing the standard deviation about the mean recession velocity for tracers with $\Delta RA > 200$ arcsec, and obtain $\sigma=30 \pm 7$ km s$^{-1}$, which is consistent with what is expected from the measured stellar mass ($\sim$~20--45 km s$^{-1}$; \citealt{Norris14}). Fitting this feature properly, along the stream trajectory in phase-space, requires a detailed model (e.g. \citealt{Sanderson13}). Such a model should return a substantially lower value for the dispersion, which would not be implausible since stream dispersions are expected to vary rapidly by factors of a few, and to decrease monotonically with time owing to conservation of phase-space density (e.g. \citealt{Helmi99, Font06}).

\subsection{Stream tracers}\label{sec:streamtracers}

In summary, we find a total of 4 GCs and 10 PNe that are likely associated with the stream, as well as the probable stream nucleus. All of the \hii\ regions were found to have kinematics consistent with the disc of NGC~4651. Disc-like kinematics were also found for the objects superimposed on the Western plume and shells (D--F), which could imply either a disc origin for these substructures, or a coincidence between disc and stream in projected coordinates. The morphologies of these features suggest that the shells E--F are stream components, while the plume D may include a combination of stream and kicked-out disc material.

As a plausibility check on the number of tracers identified with the stream, we consider specific frequency expectations: the numbers of GCs and PNe associated with a given luminosity of stellar material. Focussing on the East side where the stream photometry and tracer identifications are clear, we note that the stream in this region has a $B$-band absolute magnitude of $M_B \sim -15.8$ (using standard SDSS colour transformations), and spectroscopic confirmation of 6 associated PNe and 2 GCs. Using the specific frequency trends summarized by \citet{Coccato13}, we expect typically $\sim$~3 PNe and $\sim$~5 GCs (after correcting for the limited depth of the GC observations). As a point of comparison, the Local Group dwarf elliptical NGC~205 has a similar luminosity to the Umbrella region, and  hosts $\sim$~6~bright PNe and $\sim$~10 GCs \citep{Forbes00,Corradi05}.

Given the large galaxy-to-galaxy scatter in PN specific frequency, and the likelihood of a higher PN fraction in a relatively young stellar population, our observed PN numbers are consistent with the expectation (see also Section~\ref{sec:dist}). Furthermore, their spatial distribution in Fig.~\ref{fig:targets} appears to be roughly coincident with the umbrella morphology. The number of stream GCs so far identified is relatively low, but we have not yet observed all candidates. Moreover, it is possible that many of the GCs have been tidally stripped away in an earlier phase of the dwarf disruption due to their lower initial binding energies,  so that they now reside far away from the nucleus of the stream, along its leading and trailing arms in regions that are more difficult to correlate with the visible substructures. This binding-energy bias will be illustrated with an $N$-body simulation in Section~\ref{sec:nbody}.

We also expect a fair handful of stream GCs and PNe to be found on the West side of the galaxy, which appears to host significant components of the stellar substructure. The stream tracers in this region are not as readily distinguishable from the disc and halo populations, although we have found two very good stream candidates.

Regardless of the uncertainties in specific frequency, our net finding is that it is clearly feasible to map out the kinematics of distant and faint halo streams using discrete tracers and modest amounts of telescope time.

\section{Modelling}\label{sec:models}

To help interpret the positions and kinematics of substructures around
NGC~4651 and understand their possible origins, we model the data using test particle orbits 
(Section~\ref{sec:testparticle}) and support our findings using a rescaled $N$-body simulation
(Section~\ref{sec:nbody}).

\begin{table*}
\begin{center}
\caption{Range of parameters searched for the two sets (column 1) of the test particle orbit modeling and the values of the two well matching models displayed in Figures \ref{fig:spatial} and \ref{fig:rv}. The $z$-coordinate and -velocity ($w$) of the tentatively identified nucleus are given in columns 2 and 3, respectively. The amplitude ($v_p$) and direction ($\theta = \arccos(u/v_p)$) of the velocity component in the $x$--$y$-plane can be found in columns 4 and 5, respectively. Columns 6 presents the circular velocity ($v_c$) while  column 7 lists the core radius ($r_c$) of the potential as parameterised by equation \ref{eq:logpot}.}\label{table:MCMCparams}
\begin{tabular}{lcccccc}
\hline\hline
Set&	$z$&		$w$&	$v_p$&	$\theta$&		$v_c$&	$r_c$\\
&(kpc)&(km s$^{-1}$)&(km s$^{-1}$)&(deg)&(km s$^{-1}$)&(kpc)\\
(1)&(2)&(3)&(4)&(5)&(6)&(7)\\
\hline
\multicolumn{7}{c}{Searched Range}\\
\hline
N&	 $-45$ \dots 0&	$-75$ \dots $-30$&	50 \dots 170&	8 \dots 10&			200 \dots 230&	0 \dots 10	\\
P&	0 \dots 45&		$-75$ \dots $-30$&	$-170$ \dots $-100$& 8 \dots 10&			200 \dots 230&	0 \dots 10	\\
\hline
\multicolumn{7}{c}{Good match parameters}\\
\hline
N& $-27$ &	$-75$&	139&	8&	200&	7	\\
P&	14&		$-75$&	$-166$& 	8&	200&	7	\\
\hline
\end{tabular}
\end{center}
\end{table*}

\subsection{Test Particle Orbits}\label{sec:testparticle} 

As an initial step, we match simple test particle orbits to the stream data. This approach is analogous to that taken by \citet{Law09} for the Sagittarius Stream in the Milky Way to constrain the relevant parameter space before fitting the more realistic, but also more expensive, $N$-body models \citep{Law10}. This method has also been used in various other papers for modelling streams in the Milky Way \citep{Willet09,Newberg10,Koposov10}.  We note that while this approach cannot constrain the shape of the underlying potential \citep{Varghese11, Sanders2013, Lux13}, since even thin streams do not delineate single orbits \citep{Eyre11}, it is sufficient to confirm the dynamical history of the progenitor and to place first order constraints on the satellite's orbit. This is because missing coordinates along the stream can be derived from other coordinates without knowledge of the underlying gravitational potential
\citep{Jin2008, Binney2008, Eyre2009, Jin2009, Lux13}. However, the simple test particle orbit cannot be interpreted as the orbit of the progenitor nor of any star within the stream. It is merely a construct to estimate the peri-/apocenter and period of the orbit.
\subsubsection{Method}

We use the method presented in \cite{Lux13}, where the host galaxy potential is represented by a cored logarithmic potential
\begin{equation}
\label{eq:logpot}
        \Phi=v_c^2 \ln ( x^2 +  y^2 + z^2 + r_c^2).
\end{equation}
This function corresponds to the spherical version of the halo potential used in \citet{Law05}. In it, we integrate test particle orbits using the code {\tt Orbit\_Int} \citep{Lux10}, while neglecting any second order effects such as dynamical friction. Without loss of generality, we use the nucleus of the progenitor galaxy as a starting point for our integration to trace
the orbit forwards and backwards in time. The orbit integration has several unconstrained or unknown parameters, including some of the coordinates of the initial conditions as well as parameters of the potential. We efficiently scan this parameter space for suitable orbits using a Markov-Chain Monte-Carlo (MCMC) code that has been previously presented in \citet{Lux12, Lux13}, largely following the algorithm described in \cite{Simard02}.

A summary of all model parameters and the ranges searched by the algorithm can be found in Table \ref{table:MCMCparams}, where they are split into two sets depending on whether the nucleus is located in front of or behind NGC~4651 from the perspective of the observer. While we have clearly identified these two distinct models, due to the large parameter space we cannot exclude that other models might be equivalent or better matches to the observational data. A description of our selection of free parameter ranges is given below.
\begin{itemize}
\item We keep the ($x,y$)-position fixed by the data for object p1\_sup\_79 (i.e. the stream nucleus) given in Table \ref{table:values} (where $x$ and $y$ are equivalent to the sky positions $\Delta$RA\ and $\Delta$Dec\ relative to the galaxy centre). The angular positions in arcseconds have been converted into kpc assuming a distance of 18.7\,Mpc.
\item We vary the $z$ coordinate of the initial conditions, i.e. the coordinate perpendicular to the plane of the sky ($x$--$y$ plane) in a right-handed coordinate system. In Set N, we assume that the tentatively-identified nucleus is on the observer's side of the galaxy, and therefore can only have negative $z$ values. This will correspond to a model solution where the stream progenitor is approaching apocentre towards the observer, and moving Eastwards across the sky, so that the umbrella plume represents the leading part of the stream. The maximum $z$ value has been set to 1.5 times the maximum distance of the stream in the $x$--$y$ plane. We assume the same range of $z$-coordinates on the other side of the galaxy for Set P, where the stream progenitor is leaving apocentre and approaching the observer while it moves Westwards across the sky. For both model sets, the stream orbit moves counter to the disc rotation.
\item We scan the range of the three velocity components of the progenitor, where the velocity components in the $x$, $y$ and $z$ directions are defined as $u$, $v$ and $w$, respectively. The component $w$ may vary within $\pm1\sigma$ of the observed value\footnote{We have also performed tests using $\pm3\sigma$ and found that the results are comparable.} as given in Table \ref{table:values}. The limits of the amplitude of the velocity in the plane of the sky ($v_p$) and its orientation ($\theta$) have been chosen to align the orbit with the substructure data for each set of parameters, respectively. The components $u$ and $v$ are related to these parameters via the formulae $u=v_p \cos(\theta)$ and $v=v_p \sin(\theta)$.
\item We vary both the circular velocity ($v_c$) and the core radius ($r_c$) of the potential. The range of $v_c$ has been chosen to include the measured value of the observed,  deprojected maximum rotational velocity as measured by \citet{Epinat08}. The allowed range of $r_c$ has been chosen to cover a large range of possible values.
\end{itemize}

The observables that we match are the stream positions on the East side (B region in Fig.~\ref{fig:spatial}) and the possible continuation on the West side (C). The Western plume (D) is not directly matched, but the initial parameter range has been chosen such that its position is best reproduced. Matching all three of these features simultaneously is difficult because the $x$-values for the two Western features overlap and create an ambiguity in the matching routine. None of the tracer velocities are actively matched, but the stream progenitor velocity is implicitly matched as one of the model parameters.

\subsubsection{Resulting models and interpretation} \label{sec:modelresults}

We find two MCMC solutions with low $\chi^2$, representative of  the parameters sets N and P. As this model is not a quantitative fit to the data, $\chi^2$ values do not have the meaning typically associated with them, but have merely been used as a guidance towards the goodness of the match. The  model parameters are reported in Table~\ref{table:MCMCparams} and the solutions illustrated in the upper two panels of Fig.~\ref{fig:spatial}. The spatial positions of these models qualitatively match all of the main substructure features A--F, even though
some of these features were not explicitly matched. This match strongly supports the dynamical connection between the different parts of the stream, which cannot be seen directly in the observations owing to the presence of the main galaxy disc. Furthermore, the results contribute additional evidence that the previously tentatively-identified nucleus is the progenitor of the stream, as both the leading and trailing arms have comparable lengths.  

We were not able to match all of the features in detail, which was anticipated to be a  limitation of the simple orbit approach. It might appear that model N is preferred owing to the better match to the radial position of the Western shell F, but variations at this level are naturally expected for a realistic stream composed of a spread of stellar orbits (as will be seen in Section~\ref{sec:nbody}). It is also possible that the stream configuration is different than modelled here, such that the plume D is not part of the stream, and the second Western shell feature E may mark an alternative apocentre (or a bifurcated stream). However, the fact that a single model can reproduce the position of plume D relatively well is a strong indication that it shares a common origin. In general, our orbit modelling parameter space was by no means exhaustive and guaranteed to include the correct solution, but rather to provide some plausible possibilities.

Turning to the kinematics, the stream model tracks in Fig.~\ref{fig:spatial} are  colour-coded by their line-of-sight velocities, and plotted in position--velocity phase-space in Fig.~\ref{fig:rv}
(where models N and P are the black solid and dashed grey curves, respectively). Although there was no attempt to fit the observed kinematics, the models do reproduce nicely the kinematics of the umbrella (A,B), modulo a velocity gradient that may be stronger than in the observations and which would require more detailed modelling to address. The models also predict velocities on the West side that are similar to the disc kinematics, which does not relieve the uncertainties discussed in Section~\ref{sec:kin} about the discrimination between disc and stream material in this region.

Unfortunately, the sparseness of the kinematical data and the limitations of the single test particle approach do not permit us to distinguish between the two model scenarios N and P.  
A more detailed model, such as an $N$-body simulation, may help to resolve this uncertainty.

Some physical parameters of interest for the stream model solutions
are $r_{\rm peri} \approx 4$\,kpc, $r_{\rm apo} \approx (40\pm5)$\,kpc and $t_{\rm period} \approx (385\pm10)$\,Myr for parameter Set N. The orbit values for parameter Set P are $r_{\rm peri} \approx 1.5$\,kpc, $r_{\rm apo} \approx(35\pm5)$\,kpc and $t_{\rm period} \approx (315\pm10)$\,Myr. We infer that the stream very likely has a pericentre $r_{\rm peri} \sim$\,2--4\,kpc and an apocentre $r_{\rm apo} \sim 40$\,kpc, which are not too different from the observed, projected apocentre and pericentre distances. The orbital eccentricity is $e=(r_{\rm apo}-r_{\rm peri})/(r_{\rm apo}+r_{\rm peri}) \sim$\,0.8--0.9. The tentative value for the period $t_{\rm period} \gtrsim 300$\,Myr cannot be constrained directly from the observations: all back-of-the-envelope calculations of its value would use the same data as in the creation of this model, and hence would not add information.

To consider the stream orbit relative to the host galaxy disc, we note first that the near side of the disc is clearly to the South, based on the dust lane silhouettes on the bulge in the images. The example orbit model P is inclined $\simeq 48^\circ$ relative to the observer, and $\simeq~32^\circ$ to the disc. We remind the reader that inclinations derived from orbits in spherical potentials have large modelling uncertainties as orbital precession is assumed not to exist. The orbit passes through the disc only near pericentre 
(at radii of $\sim$~2~kpc). 
Example model N is closer to an edge-on inclination ($\simeq 79^\circ$) and follows a quasi-polar orbit ($\simeq 54^\circ$). It has  
three disc passages, near both pericentre and apocentre 
($\simeq$\, 7, 17 and 34 kpc).

The latter solution is preferred in order to explain the strong disturbances in the \hi\ and stellar discs at 
$\sim$~20--35~kpc.
Its leading arm crosses the disc on the West side in the Westward direction, which would make sense for the direction of plume D if it includes kicked disc material. The impact velocities when the nucleus crosses the disc are $\sim$\,500, $\sim$\,350 and $\sim$\,200\,~km~s$^{-1}$ from the inner to outer regions.

In summary, the combination of fairly complete positional information for multiple stream features, along with a single kinematic data point, has allowed us to define the stream orbit remarkably well. In fact, the solution is limited not by the lack of constraints but by the limitation of a single-orbit model in representing an extended stream.

As pointed out before, single test particle orbits cannot accurately model streams due to the stream-orbit offset. Before continuing in the next Section to examine an $N$-body simulation, we consider some analytic expectations for this deviation of a stream from a single test particle orbit, considering the large spread of orbital energies of individual stars in the stream. For debris generated from objects disrupting along mildly eccentric orbits with galactocentric radius $r$, both the fractional width and height of a stream ($h/r$), and its fractional offset from its orbit ($\Delta r/r$) should be of similar order to the tidal scale,
\begin{equation}
	\frac{h}{r} \sim \frac{\Delta r}{r} \sim \left({m_{\rm sat} \over M_{\rm Gal}}\right)^{1/3}
\label{eqn:streamwidth}		
\end{equation}
\citep{Johnston98,Johnston01}, with a mild dependence on time becoming important after multiple orbits \citep{Helmi99}. For highly eccentric orbits such as the one seen here, the expected stream width and orbital offset would be much higher. 

For the umbrella system, we estimate 
$r\simeq$~ 35~kpc 
at the current stream nucleus position, $m_{\rm sat}\sim 6\times10^8 M_\odot$ for the stream stellar component, and $M_{\rm Gal} \simeq 5\times10^{11} M_\odot$ for the total host mass within 
35~kpc.
Therefore in the low-eccentricity limit, we expect a stream height and offset of 
$h \sim \Delta r \sim$~ 3.5~kpc. This value is intermediate to the 
$\sim$~200~pc  projected width of the narrowest part of the stream, and the 
$\sim$~20~kpc  spread of the umbrella shell.

\subsection{Rescaled $N$-body Simulation}\label{sec:nbody}

\begin{figure*}
\begin{center}
\includegraphics[width=88mm]{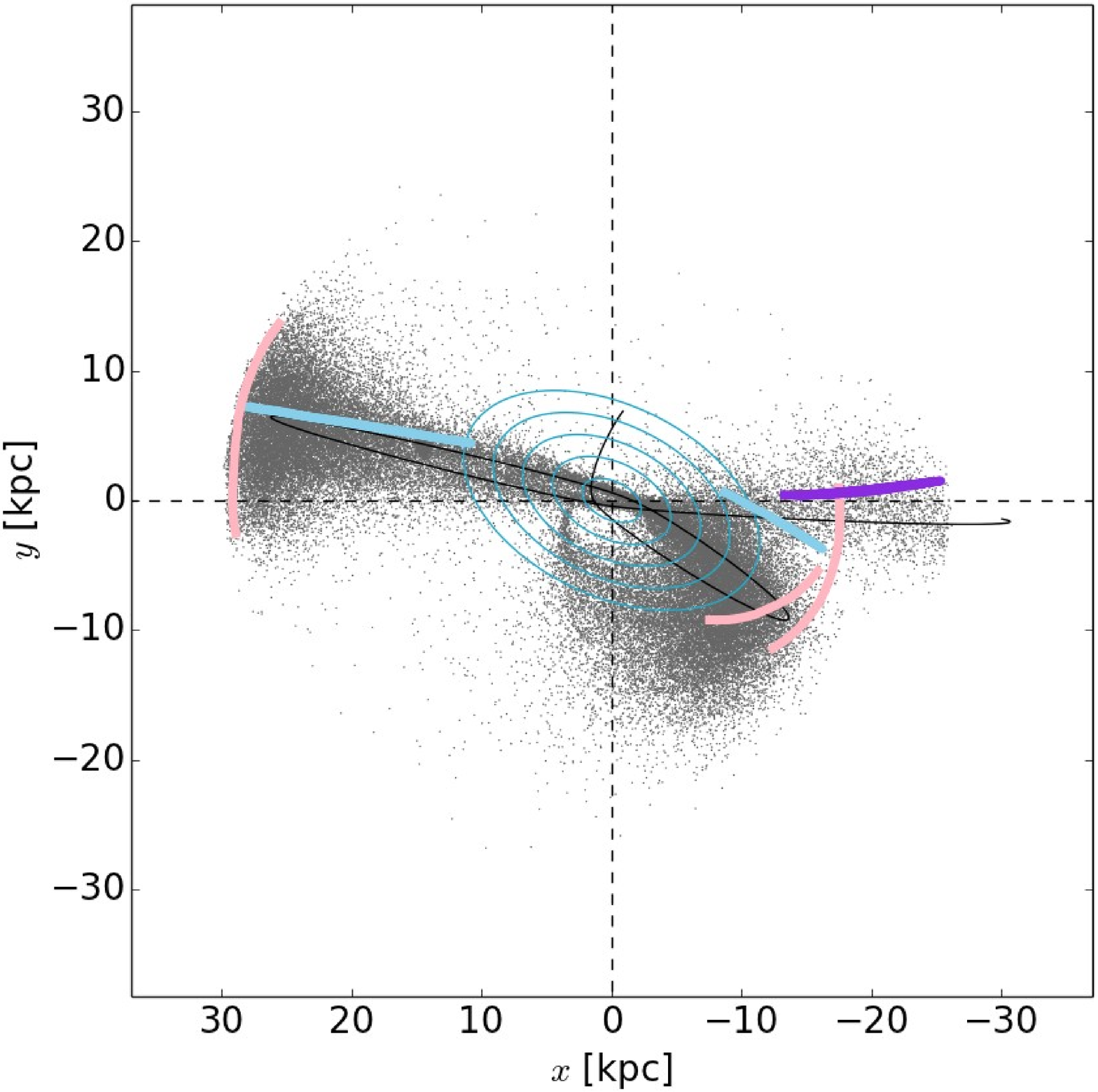}
\includegraphics[width=88mm]{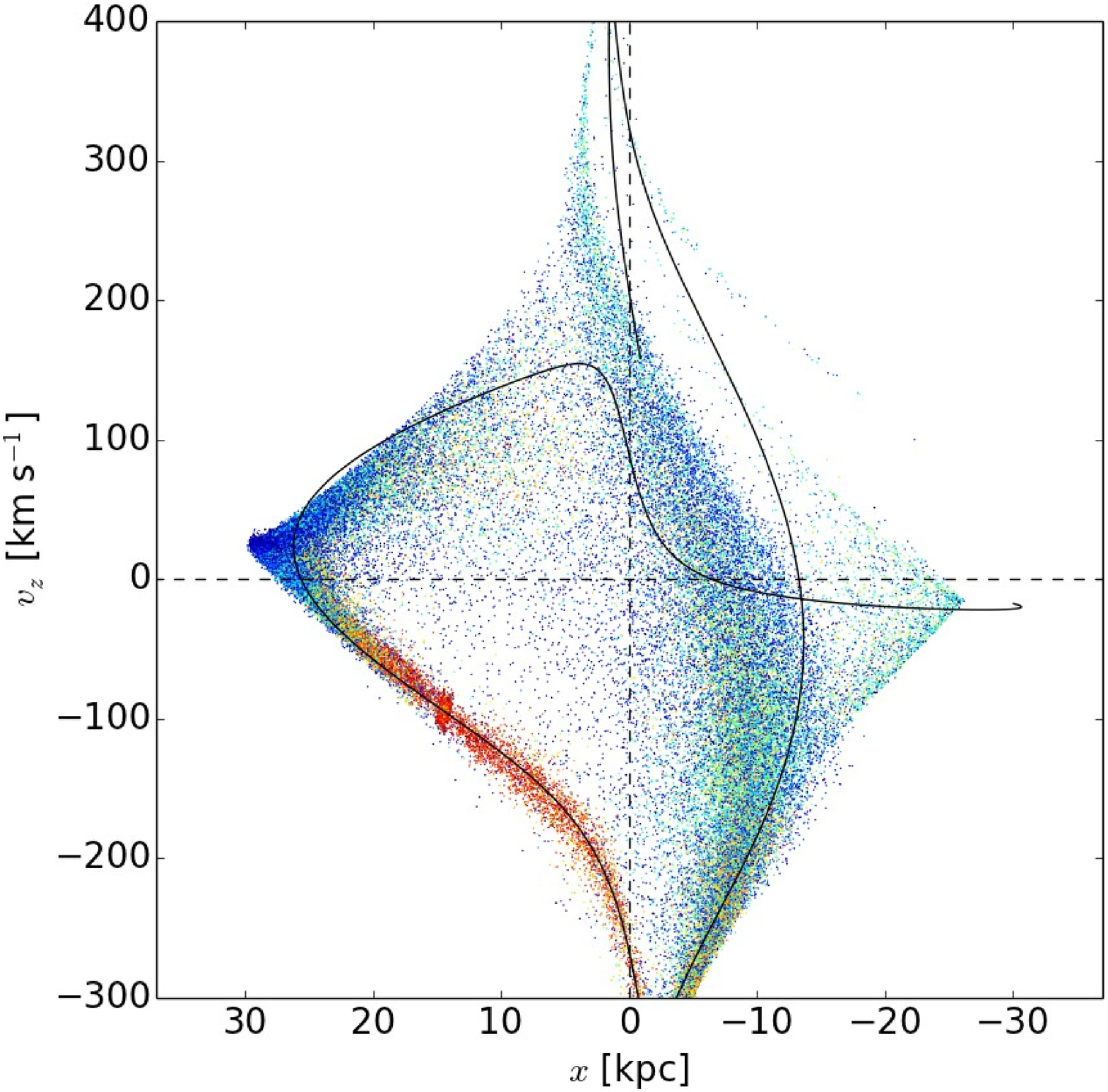}
\includegraphics[width=88mm]{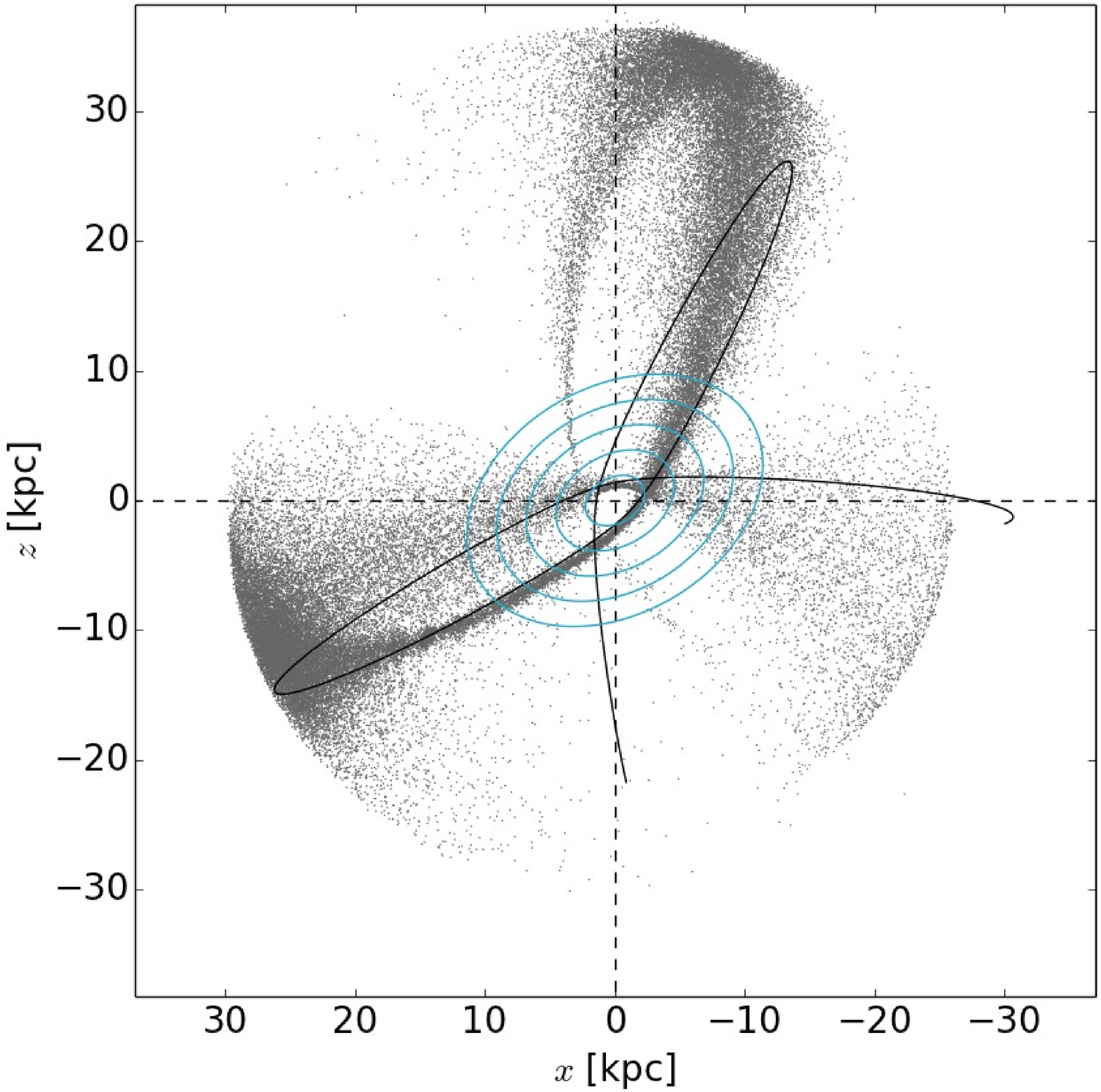}
\includegraphics[width=88mm]{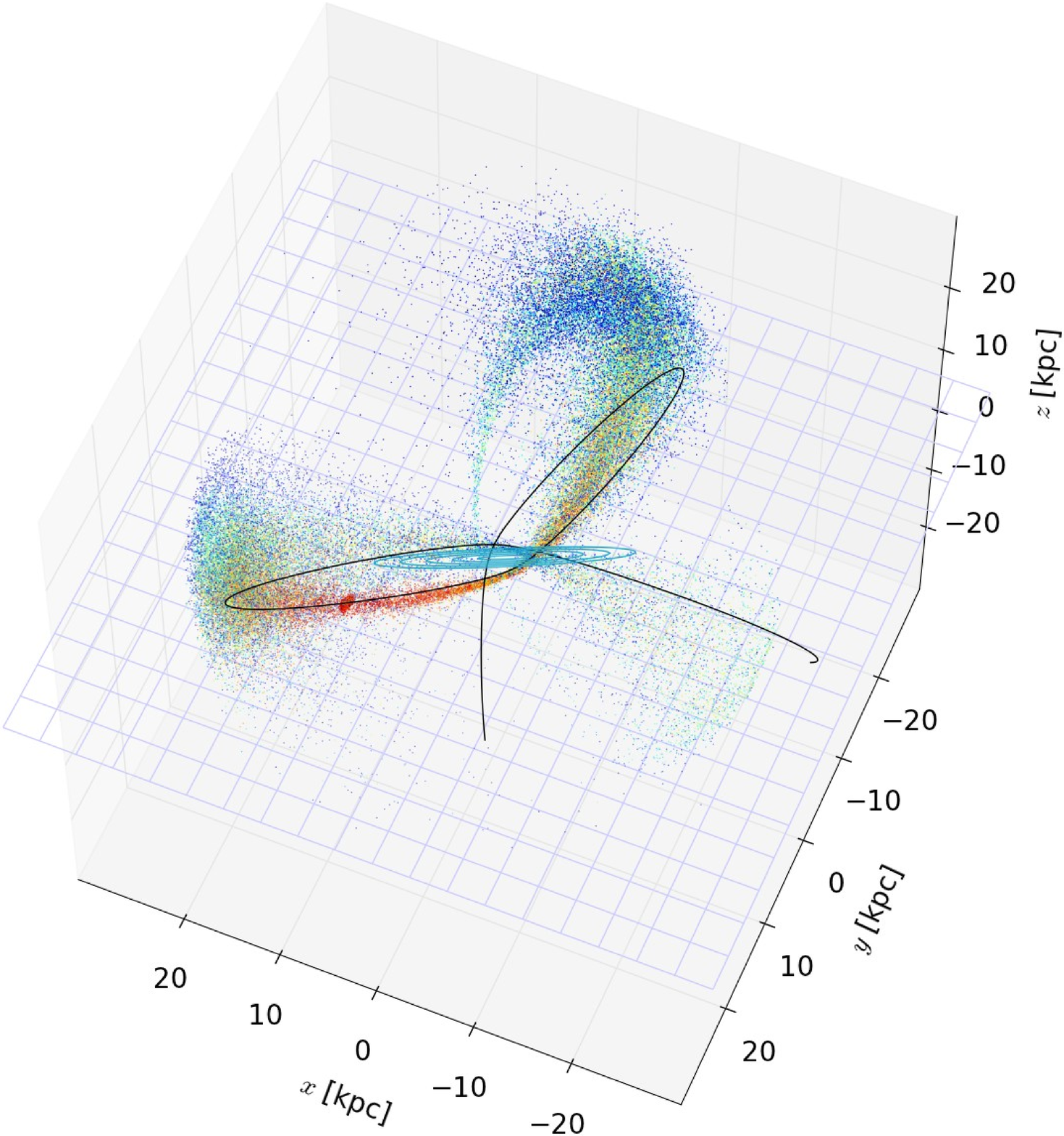}
  \caption{$N$-body model of an Umbrella-type stream, where the small dots correspond to the $N$-body particles. The black curve represents the inferred orbit of the progenitor (single particle orbit model `N'), and the light blue ellipses show the stellar disc in spacings of one disc scale-length. The upper-left panel shows the $x,y$ positions on the sky, with the coloured curves corresponding to the features labeled A--F in Fig.~\ref{fig:spatial}. The upper-right panel shows a position--velocity phase-space diagram (cf.\ Fig.~\ref{fig:rv}), where the particles are colour-coded by initial binding energy within the progenitor satellite (red for most bound, blue for least bound). The lower-left panel is an orthogonal projection to the sky so that line-of-sight distances $z$ are visible; the remnant of the progenitor nucleus is seen as a clump on the near side of the galaxy, which is approaching apocentre. The lower-right panel shows a tilted view to illustrate the trajectory of the stream relative to the disc; the stream passes through the galaxy at small radii on one side, and at a large radius on the other. An interactive 3D pdf version of this figure is available online as supplementary material.}
\label{fig:Nbodyspatial}
\end{center}
\end{figure*}

\begin{table*}
\begin{center}
  \caption{Comparison of original and rescaled parameters ($\gamma=0.36$) of the $N$-body simulation against observational constraints. The mass ($M_{\rm host}$), virial ($r_{\rm vir}$) and scale ($r_s$) radii as well as circular velocity ($v_c$) of the simulated host potential are given in columns 2 to 5, respectively. The host potential is modelled using an NFW profile combined with a central Hernquist stellar profile. The mass ($M_{\rm sat}$), effective radius ($r_e$) and velocity dispersion of the satellite ($\sigma_{\rm sat}$) are given in columns 6, 7 and 8, respectively. Columns 9, 10 and 11 list its initial apo- ($r_{\rm apo}$) and peri- ($r_{\rm peri}$) centre radii, and the stream's expected disruption time scale ($t_{\rm d}$).
}\label{table:rescale}
\begin{tabular}{lcccccccccc}
\hline\hline
set & $M_{\rm host}$ & $r_{\rm vir}$ & $r_s$ & $v_c$ & $M_{\rm sat}$ & $r_e$ & $\sigma_{\rm sat}(r_e)$ & $r_{\rm apo}$ & $r_{\rm peri}$ & $t_{\rm d}$ \\
 & ($M_\odot$) & (kpc) & (kpc) & (km s$^{-1}$) & ($M_\odot$) & (kpc) & (km s$^{-1}$) & (kpc) & (kpc) & (Myr) \\
(1) & (2) & (3) & (4) & (5) & (6) & (7) & (8) & (9) & (10 ) & (11) \\
\hline
original&$4.2\times10^{14}$&	$1550$&	564&475&		$1.8\times10^9$&		1&				35&					90&			2&	500\\
rescaled&	$2.0\times10^{13}$&	$558$&		203&		171&			$8.4\times10^7$&		0.4&				13&			32&	0.7&	500\\
observed& -&				-&				-&			215$^*$&			$3.4\times10^{8\dagger}$&			-&				-&					$\gtrsim 36^\dagger$& -&-\\
\hline
\end{tabular}
\end{center}
References: $^*$ \cite{Epinat08}, $^\dagger$ this work
\end{table*}

While simple test particle orbits are not sufficient to properly model the kinematic data of this massive stream, the more accurate $N$-body models are very expensive to calculate, and extensive parameter searches that explore all possible degeneracies that can arise from, for example, internal rotation \citep{Penarrubia2010} are not yet computationally feasible. Here, we instead use a single $N$-body simulation  to support our previous results and qualitatively motivate the dynamical connection between the different arms of the stream, as well as the identification of the stream progenitor nucleus.

We use a simulation first described in \citet{Romanowsky12} that has been modelled in a rigid, spherical host-galaxy potential, comprising combined \citet{Hernquist1990} and \cite{Navarro1996} models for the stellar and dark matter contributions. The satellite galaxy was on a very eccentric orbit, and modelled as a self-gravitating $N$-body system, with no differentiation between stellar and dark matter particles, and was integrated using a tree code \citep{Hernquist1987}. 

The original parameters of the simulation are summarised in Table \ref{table:rescale} together with the observed data and the correspondingly rescaled simulation. For the rescaling, we follow \cite{Helmi03} and \cite{Holopainen06}, who kept the density and therefore the dynamical time of the simulation constant in the process. This results in a cubic scaling for the masses as $M_n = M_o \gamma^3$, while distances and velocities scale linearly $r_n = r_o \gamma$, $v_n = v_o \gamma$. However, because the simulation was not originally designed for NGC~4651, e.g. with an appropriate choice for apocentre, there is no single rescaling that matches all of the observables. Rescaling to match the circular velocity of the host galaxy would mean a scaling factor of $\gamma=0.45$, while rescaling to match the radii of the shells would mean $\gamma \simeq 0.36$. We adopt the latter value, since our interest here is in reproducing approximately the positions and morphologies of the substructures on the sky, with the consequence that the values of the velocities in this model cannot be compared directly to the data.

After rescaling, we have selected viewing angles that provide a reasonable representation of the observations. The fact that we are able to do so at all -- for a simulation that was generated for a completely different set of observations -- is a powerful reminder of the generality of tidal stream morphologies and dynamics. The upper-left panel of Fig.~\ref{fig:Nbodyspatial} shows a projection on the sky of the $N$-body particles from the simulation (dots) and the orbit of the progenitor (black curve), compared to the observed features of the umbrella stream (coloured lines).  The orbit configuration relative to the observer is similar to example model N from the previous Section. Even though the simulation was not purpose built, it provides a good match to the data and qualitatively reproduces the observed shell features.

Comparing the paths of the stream and of the nucleus, one can see offsets in both angle and radius  owing to the internal spread and self-interaction of the satellite. This is a reminder of the approximate nature of  single-particle orbit modelling, but also a confirmation that the approach can still provide valuable constraints -- in this case the apocentre and pericentre of the orbit as well as its dynamical time to within 10\%.

In the upper-right panel we show the corresponding 1D phase space plot. It shows the characteristic chevron-shaped behaviour of a shell, with particles around its edge lying near apocenter with very low velocities, which increase in a nearly straight line in projection as the particles approach pericenter \citep{Merrifield98}. More detailed modelling of this phase-space trajectory could be used to provide an independent estimate of the gravitational
potential of the host galaxy \citep{Sanderson13}.

The particles in the phase-space panel are also colour-coded by initial binding energy within the satellite galaxy, with red being the most bound. It can be seen that the particles have retained a strong memory of their initial conditions, with the least bound material found far away from the progenitor nucleus. Therefore if the progenitor had any internal stellar population gradient, this could be manifested observationally as a systematic colour variation along the stream (which we have so far not seen in the Umbrella system).

The lower panels show additional projections of the simulation. The stream orbit has an orientation between polar and equatorial relative to the disc ($\sim 58^\circ$). The stream passes through the inner disc three times, at radii of $\sim$~1.5--5~kpc. It also passes at a low angle through the outer disc on the East side, at a radius of $\sim$~14~kpc. For additional visualization, an interactive 3D plot is provided as Supplementary Material.

\section{Discussion}\label{sec:discussion}

We next place the Umbrella galaxy in observational and theoretical context, comparing it to some other well-known streams (Section~\ref{sec:comp}), examining how (a)typical an event it represents in a cosmological context (Section~\ref{sec:stat}), and considering some implications for the effects of minor mergers on host galaxy discs (Section~\ref{sec:dyn}).

\subsection{Stream comparisons}\label{sec:comp}

We first make comparisons with two famous substructures in the Local Group: the Sagittarius stream around the Milky Way (Sgr; \citealt{Ibata94,Dolphin02,Majewski03,Chou07,NiedersteOstholt10}), and the Giant Southern Stream around M31 (GSS; \citealt{Ibata01b,Fardal06,Tanaka10}). These all have similar luminosities of $\sim 10^8 L_\odot$, and orbit around spiral galaxies with circular velocities of $\sim$~200~km~s$^{-1}$ and stellar masses of $\sim5\times 10^{10} M_\odot$, implying stellar mass ratios of $\mu_\star \sim 0.02$. The stream progenitors would have all been gas-poor dwarf galaxies, perhaps similar to NGC~147, NGC~185, and NGC~205 (satellites of M31).
However, the Umbrella Galaxy may be younger and more metal-poor than the other two ([Fe/H]~$\sim -0.9$ vs.\ $\gtrsim -0.6$, and $\sim$~4~Gyr vs.\ $\sim$~8--9 Gyr, see Section \ref{sec:images}).

\begin{figure*}
\begin{center}
\includegraphics[width=195mm]{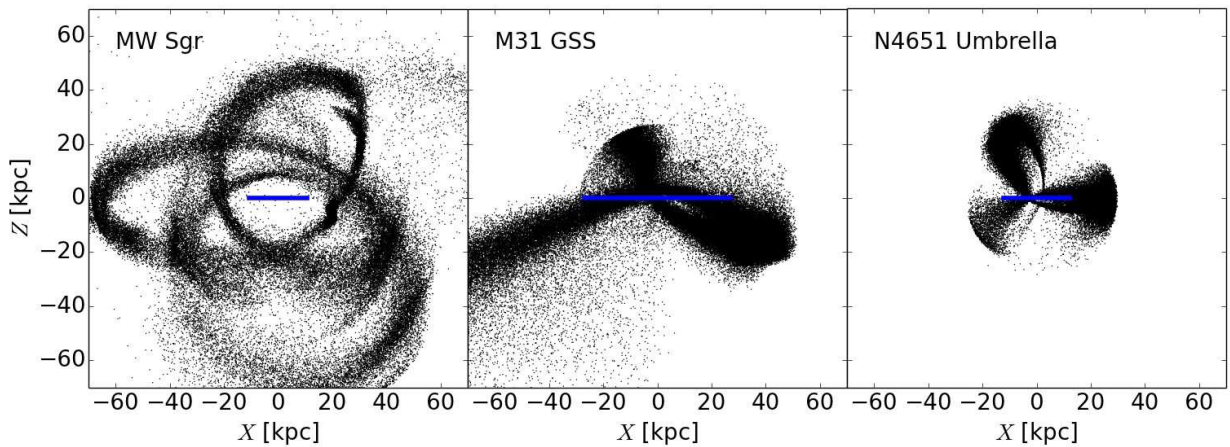}
\caption{ $N$-body realizations of three streams in the local Universe. The $X$ and $Z$ coordinates are in the host galaxy reference frame, with the edge-on discs seen as blue lines out to 5 disc scale-lengths. Although all three streams are similar in terms of stellar masses and progenitor types, the orbits are very different, with the GSS and the Umbrella being much more eccentric than the Sgr stream, and consequently producing more shell-like morphologies.
}
\label{fig:streams}
\end{center}
\end{figure*}

The orbits of these streams may be visually compared in Fig.~\ref{fig:streams}, where $N$-body realizations are plotted (the Sgr and GSS models taken from \citealt{Law10} and \citealt{Fardal12}, respectively). The GSS and Umbrella are morphologically similar, with highly eccentric orbits that produce shells, and pericentre and apocentre distances of a few kpc and $\sim$~30--40~kpc, respectively. The Sgr stream is less eccentric, producing more classically stream-like features rather than shells, and orbits much further out ($\sim$~20--60~kpc). The Umbrella has the most compact orbit, and consequently the highest surface brightness. All three streams have orbits that are fairly polar, with their angular momentum vectors roughly orthogonal to the spin axes of their host galaxy discs.

In some regards, the stream around the Umbrella Galaxy is an analogue to both Local Group streams, which could provide complementary information about this type of event.  Although much more detail can be obtained for Local streams, such as 6D position--velocity information, the greater distance of the Umbrella Galaxy also affords some advantages. The entire system can be observed within a single telescope pointing, and the use of integrated light rather than resolved star counts permits the stream morphology to be more readily discerned.

\subsection{Stream statistics and comparisons to theory}\label{sec:stat}

These streams are not only interesting because of their spectacular nature, but also important because they may represent a significant pathway in the stellar mass growth of luminous spiral galaxies -- contributing in particular to their stellar haloes and GC systems. The Sgr stream is thought to have deposited $\sim$~6 GCs \citep{Geisler07}, and its stellar mass
is comparable to the total stellar halo mass of the Milky Way \citep{Bochanski14}, while the GSS represents a smaller fraction of the M31 stellar halo (which is more massive than that of the Milky Way; \citealt{Ibata14}). In NGC~4651, there are only a handful of halo PN candidates besides those associated with the stream, which suggests that the Umbrella stream will comprise a large fraction of the diffuse stellar halo of this galaxy, once it is phase-mixed.

All three of these streams appear to represent dominant or even unique events in the assembly of their host galaxy stellar haloes, which raises some interesting questions about their typicality. Are these galaxies representative of low-redshift spirals -- in other words, are the observed $\mu_\star \sim 0.02$ accretion events typical for recent epochs? And beyond the qualitative consonance of these accretion events with the hierarchical formation paradigm, do the stream demographics mesh quantitatively with theoretical expections?

Answering these questions for galaxies beyond the Local Group is difficult using the traditional technique of sifting the stellar haloes for mixed-in signatures of past events (although see \citealt{Mouhcine10,Tanaka11,Greggio14}). Alternatively, one may search for cases of well-defined substructures that trace ongoing accretion events, and combine their frequency of occurrence with a model for their visibility times to arrive at an estimate for the event rate in the overall galaxy population.

Such work is far beyond the scope of the present paper, but we can usefully sketch out what is known to date, in order to get a rough sense for the context of the three streams. First, the observed frequency of clear halo substructures around low-redshift spiral galaxies is $\sim$~5\% \citep{Miskolczi11,Atkinson13}. Such features are expected to persist for a few orbital time-scales \citep[e.g.][]{Rudick09,Gomez13}, which would be $\sim$~1~Gyr if the Umbrella system is representative. The implied frequency of accretion events at $z\sim0$ would then be $\sim$~0.05~Gyr$^{-1}$, meaning the Umbrella is {\it not} a typical event for low-redshift spirals. However, if the accretion rate was higher at earlier times (as expected for an expanding universe), then it becomes plausible that at some point in their histories, virtually all spiral galaxies experienced an event like the Umbrella, with a stellar mass ratio of $\mu_\star \sim 0.02$.

The observed frequency of stellar streams can be compared directly to theoretical predictions only once a large enough sample of galaxies is simulated in order to average out the stochasticity of accretion histories (cf.\ \citealt{Cooper10}). However, indirect comparisons can already be made if one makes reasonable assumptions about the total masses (including dark matter) of the infalling satellites. The standard approach here is to use the result of abundance matching in a $\Lambda$CDM cosmology, where the observed luminosity function of galaxies is combined with the predicted halo mass function to give the implied stellar-to-total mass ratio. For the Umbrella Galaxy system, the implied total masses of the main galaxy and the satellite are $\sim 6\times10^{11} M_\odot$ and $\sim 9\times10^{10} M_\odot$, respectively \citep{Behroozi13}. The implied total mass ratio of $\mu_{\rm tot} \sim 0.15$ (from a combination of observed $\mu_\star$ and predicted halo mass) is similar to the ``dominant'' mode of accretion for Milky-Way mass galaxies in $\Lambda$CDM of $\mu_{\rm tot}\sim0.1$, with such events occurring today with a frequency of $\sim$~0.05~Gyr$^{-1}$ \citep{Stewart08,Stewart09b}.

This frequency meshes nicely with the empirical rate of prominent substructures in the nearby universe, as calculated very roughly above, and suggests consonance between theory and observation. The overall implication is that the Umbrella stream may represent a pivotal pathway for galaxy assembly in a $\Lambda$CDM context. One caveat here is that recent semi-analytic models have predicted the most massive progenitor of the stellar halo in a Milky Way-mass galaxy to have $M_\star \sim 5\times10^7 M_\odot$, with $M_\star \sim 4\times10^8 M_\odot$ as in the Umbrella being very unusual \citep{Cooper13}. There may also be tension between the implied total mass ratio and the disk dynamics (see next Section).

Beyond the masses of accretion events and streams, it may be informative to compare the observed orbital elements (turning points and angular momentum direction) with predictions for cosmological infall. Interestingly, the characteristic radii of $\sim$~30~kpc of the three streams discussed above are somewhat smaller than the more typical $\sim$~50~kpc for recent satellite disruption in $\Lambda$CDM models \citep{Bullock05}. However, it is likely that such relatively massive satellites were affected more by dynamical friction and migrated to smaller radii (cf.\ \citealt{Leaman13,Chakrabarti14}).

The orbital directions of streams are relevant to questions about how galaxies acquire their angular momenta, and to controversies about the observed coherence of satellite systems (e.g., \citealt{Kroupa10,Deason11b,Bett12,Ibata13a}). Stream observations could provide unique contributions in this context, since their individual orbits can be defined much more precisely than for intact satellites. It is curious that the three streams discussed here all have fairly polar orbits, and more cases should be studied to establish the statistics of their orientations -- while being attentive to the potential for observational selection biases and to the fundamental limitations in making inferences about the original orbits at infall \citep{Lux10}.

\subsection{Stream dynamics}\label{sec:dyn}

The above schematic analysis of observed frequency and theoretical prediction seems to hang together nicely, and raises an additional point of interest. A mass ratio of $\mu_{\rm tot} \sim 0.15$ represents a dynamically significant event for the main galaxy, and therefore we might expect the umbrella stream to perturb strongly the disc of NGC~4651. \citet{Purcell11} recognized this point for the Sgr stream, which they argued to be the driving force for the spiral structure of the Milky Way. The GSS has been proposed as the cause of the warp and kicked-up disc stars in M31 \citep{Sadoun14,Dorman13}. More generally, it has long been suspected that satellite perturbations could account for many of the spiral arms and warps in disc galaxies \citep{Kormendy79,GarciaRuiz02,Kazantzidis09}.

On the other hand, there has been a concern that cosmologically common $\mu_{\rm tot} \sim 0.1$ events would be so energetic that they would thicken discs and form bulges in excess of observational constraints (e.g. \citealt{Stewart08,Purcell09,Kormendy10}). Indeed, such 1:10 mass ratio events are commonly studied within the context of transforming late-type galaxies to early-types (e.g. \citealt{Bournaud05}), where they are considered not as mere transient perturbations to the initial disc, but as a source of severe heating where even one event would lead to an S0 galaxy. This concern is a variation of the substructure crisis, whose most extreme solution is the abolition of CDM.

In this context, the Umbrella Galaxy system could provide novel insights. Given the basic properties and orbit of the stream that we have determined, is the noticeable but mild back-reaction on the host galaxy (warping and disruption of the outer stellar and \hi\ disc) consistent with  a massive satellite? This question could be addressed in part through semi-analytic models of the interaction (e.g. \citealt{Chakrabarti11, Kannan12, Feldmann13}), and ultimately will require a full hydrodynamical simulation of host and satellite on a cosmological infall orbit (cf.\ \citealt{Sadoun14}).

In the meantime, we can carry out a back-of-the-envelope calculation, following \citet{Mori08}. This is an argument from energetics that the work from dynamical friction on the satellite galaxy during its pericentric passage would be deposited as kinetic energy into the host galaxy disc. The following formula estimates the satellite mass $M_{\rm s}$ that would boost the disc scale-height by $\Delta z_{\rm d}$ :
\begin{equation}
M_{\rm s } = \left[ M_{\rm d} \, v^2_{\rm s} \, \Delta z_{\rm d} / (4 G)\right]^{1/2} ,
\end{equation}
where $M_{\rm d}$ is the disc mass and $v_{\rm s}$ is the satellite velocity. Here we note the somewhat counter-intuitive implication that faster pericentric approaches have {\it less} of an effect on the disc.

For the Umbrella system, the known parameters are $M_{\rm d} \simeq 10^{10} M_\odot$ and $v_{\rm s} \sim$~300--500~km~s$^{-1}$. Next, although we do not know the initial or final scale-height of the disc, we may adopt $\Delta z_{\rm d} \sim$~0.3~kpc as a rough maximum disturbance before it no longer resembles a normal spiral disc. We then expect that the satellite has a mass of less than  $M_{\rm s}  \sim 6 \times 10^9 M_\odot$, which is a limit that is much higher than the estimated stellar mass of the stream ($\sim 4 \times 10^8 M_\odot$), and much lower than the expected total mass of the stream progenitor ($\sim 10^{11} M_\odot$).

Thus, the observed mild reaction of the Umbrella host galaxy to the stream is plausible if the stream has lost the vast majority of its (dark) mass by the time of impact. Although this is the typical assumption in studies of stellar streams, its generic validity is by no means clear. For example, applying the calculation above to M31 and the GSS gives a progenitor satellite mass limit of $\sim 1.5\times 10^{10} M_\odot$, which in combination with abundance matching would require that $\sim$~90\% of the total mass was lost before interacting with the disc. However, the detailed simulations of \citet{Sadoun14} found that with a highly radial orbit, only $\sim$~50\% of the satellite mass within 20~kpc was lost by the time of first pericentric passage. Furthermore, an initial halo mass was adopted that was a factor of 5 lower than typical $\Lambda$CDM-based expectations, precisely because heavier haloes were found to cause too much disc damage (R. Sadoun, private communication). It would be valuable to examine this topic further for the Umbrella galaxy, with more detailed models and simulations.

\section{Summary \& Conclusions}\label{sec:conclusions}

Following the detection of a dramatic system of streams and shells around the Umbrella Galaxy (NGC~4651) by \citet{MartinezDelgado10}, we obtain 
Subaru/Suprime-Cam
 images in the standard $g'$-, $r'$- and $i'$-bands along with a narrow-band \oiii\ image.

We carry out a careful analysis of the Suprime-Cam images in order to extract the salient physiological characteristics of the substructures. Using photometry of point sources around NGC~4651, we identify candidate globular clusters, \hii\ regions and planetary nebulae. Candidates are then selected for spectroscopic follow-up with Keck/DEIMOS. A clean sample of 14 globular clusters, 19 planetary nebulae and 11 \hii\ regions belonging to the NGC~4651 system is identified based on their recession velocities. Of these, 4 globular clusters and 10 planetary nebulae are found to be associated with the faint stellar substructures.

We use the \oiii\ photometry of our planetary nebula candidates  to measure the planetary nebula luminosity function and obtain a significantly improved estimate of the distance to the Umbrella Galaxy ($19\pm1$ Mpc). This value also shows good agreement with that measured based on our inferred globular clusters luminosity function and globular cluster sizes on archival {\it Hubble Space Telescope} imaging.

Using our combined imaging and spectroscopy, we confirm that the stellar substructures are consistent with the dry accretion of a relatively bright ($M_V\sim -17.6$) dwarf satellite onto NGC~4651 in the last few Gyr.
We also tentatively identify the residual nucleus of the infalling dwarf galaxy through its location, colour and recession velocity. Integrating the light in the stream, we estimate that the stellar mass contained in the stream is at least $4\times10^8 M_\odot$. This implies a significant merger stellar mass ratio of $\mu_*\sim$ 0.02 and a total mass ratio of $\mu_{\rm tot}\sim0.15$. This mass ratio is akin to that of the Giant Southern Stream and Sagittarius Dwarf around M31 and the Milky Way systems, respectively.

Starting from the position of the tentatively identified nucleus in phase space, we run simple test particle orbit models forward and backward in order to match the positions of the various substructures. Using a Markov-Chain Monte-Carlo technique, we identify two model representations of the substructures and confirm that the identified nucleus is the likely progenitor of the stream. The infalling satellite is on a very eccentric orbit with a period of $\gtrsim300$ Myr and turning points at a few and 40 kpc. The preferred model implies a recent passage of the satellite through the disc of the host, which may explain the apparent kicked-up material (stars, ionized gas and H~{\sc i} distribution) to the West of the system.

As simple test particle orbits are not sufficient to match the range of orbits of the kinematic tracers, we also use a rescaled $N$-body simulation with similar morphology to confirm the correctness of the test particle results. At present, probing the full parameter space with $N$-body models is too computationally expensive to be feasible and a proper analysis of the stream kinematics has to be referred to future work, but it is heartening and instructive that it is possible to learn so much from a suitably ``recycled'' simulation.

This work demonstrates that it is possible to obtain spectroscopic information about faint streams using bright tracers such as globular clusters, planetary nebulae and \hii\ regions. Although we have not identified any \hii\ regions associated with this particular stream, this does not preclude the possibility that there could be associated \hii\ regions in other systems, and we have established a methodology by which such objects could be found, also providing a new method to study star formation in minor merging systems. This combination of object identification, kinematic measurement, and numerical modelling offers a powerful tool for analyzing the properties of minor mergers as they occur, thus avoiding the ambiguities of the more usual statistical analysis of systems that might merge in future. Since such mergers are believed to play a major role in galaxy formation, being able to make {\it in situ} measurements of the properties of the infalling prey, and its impact on the predator, offers a very direct way to study the evolving galaxy ecosystem.

\begin{table*}
\begin{center}
\caption{Summary of measured kinematic tracers. Column 1 gives the object ID, columns 2 and 3, its position in equatorial coordinates (J2000) and column 4 shows its heliocentric corrected recession velocity. The foreground extinction corrected $g'$, $r'$ and $i'$ and \oiii\ photometry is given in columns 5 to 8, respectively. The last column 9 shows whether or not the kinematics are consistent with the disc of NGC~4651 (see text).}\label{table:values}
\begin{tabular}{ccccccccc}
\hline\hline
ID&$\alpha$&$\delta$&$V_{{\rm obs}}$&$g'$&$r'$&$i'$&$m(5007)$&Disc\\
&(hh:mm:ss)&(dd:mm:ss)&(km s$^{-1}$)&(mag)&(mag)&(mag)&(mag)&\\
(1)&(2)&(3)&(4)&(5)&(6)&(7)&(8)&(9)\\
\hline
&&&&Nuclei&&&&\\
\hline
NGC4651\_nuc         &12:43:42.65&16:23:36.0& 795$\pm$  5& ---& ---& ---& ---&Yes\\
p1\_sup\_79      &12:43:59.27&16:24:44.8& 742$\pm$ 22&21.60$\pm$.01&21.11$\pm$.01&20.86$\pm$.01&24.0$\pm$.1&No \\
&&&712$\pm$14\textsuperscript{*}&&&&&\\
\hline
&&&&PNe&&&&\\
\hline
p1\_PN\_5        &12:43:18.03&16:24:20.9& 789$\pm$  9& ---& ---& ---&27.3$\pm$.1&No \\
p1\_PN\_7        &12:44:06.24&16:23:44.8& 778$\pm$  9& ---& ---& ---&28.0$\pm$.1&No \\
p1\_PN\_8        &12:44:03.94&16:23:55.2& 772$\pm$  8& ---& ---& ---&27.8$\pm$.1&No \\
p1\_PN\_9        &12:44:03.67&16:23:44.1& 818$\pm$  4& ---& ---& ---&26.7$\pm$.1&No \\
p1\_PN\_10       &12:44:02.57&16:25:11.2& 751$\pm$ 11& ---& ---& ---&27.4$\pm$.1&No \\
p1\_PN\_12       &12:44:02.05&16:23:46.7& 753$\pm$  6& ---& ---& ---&27.6$\pm$.1&No \\
p1\_PN\_13       &12:43:23.61&16:24:05.6& 957$\pm$  9& ---& ---& ---&27.8$\pm$.1&Yes\\
p1\_PN\_14       &12:43:57.66&16:25:15.4& 765$\pm$  7& ---& ---& ---&27.1$\pm$.1&No \\
p1\_PN\_15       &12:43:27.97&16:23:37.9& 971$\pm$  7& ---& ---& ---&27.4$\pm$.1&Yes\\
p1\_PN\_20       &12:43:31.96&16:23:02.1& 983$\pm$  9& ---& ---& ---&28.0$\pm$.1&Yes\\
p1\_PN\_22       &12:43:53.29&16:23:05.4& 707$\pm$  7& ---& ---& ---&27.3$\pm$.1&Yes\\
p1\_PN\_24       &12:43:32.31&16:23:58.4&1031$\pm$  8& ---& ---& ---&27.5$\pm$.1&No \\
p1\_PN\_25       &12:43:49.43&16:25:34.6& 703$\pm$  9& ---& ---& ---&27.6$\pm$.1&Yes\\
p1\_PN\_26       &12:43:48.44&16:25:36.9& 710$\pm$ 10& ---& ---& ---&28.1$\pm$.1&Yes\\
p1\_PN\_27       &12:43:49.94&16:25:14.5& 757$\pm$  8& ---& ---& ---&27.3$\pm$.1&No \\
p1\_PN\_28       &12:43:33.45&16:24:15.8&1053$\pm$  7& ---& ---& ---&27.8$\pm$.1&No \\
p1\_PN\_30       &12:43:37.18&16:21:58.0& 867$\pm$  9& ---& ---& ---&27.8$\pm$.1&Yes\\
p1\_PN\_31       &12:43:50.17&16:22:39.8& 719$\pm$  9& ---& ---& ---&27.7$\pm$.1&Yes\\
p1\_PN\_37       &12:43:31.10&16:23:16.3&1007$\pm$  8& ---& ---& ---& \textsuperscript{**}&Yes\\
\hline
&&&&\hii\ regions&&&&\\
\hline
p2\_HII\_5       &12:43:30.08&16:22:50.2& 948$\pm$ 20&23.65$\pm$.02&23.61$\pm$.02&24.47$\pm$.05&24.9$\pm$.1&Yes\\
p2\_HII\_6       &12:43:41.17&16:24:37.9& 815$\pm$  5&22.44$\pm$.07&22.48$\pm$.06&23.74$\pm$.22&23.1$\pm$.1&Yes\\
p2\_HII\_12      &12:43:45.85&16:22:47.2& 780$\pm$  9&23.40$\pm$.05&23.49$\pm$.05&23.86$\pm$.08&24.4$\pm$.1&Yes\\
p2\_HII\_15      &12:43:52.04&16:23:21.7& 653$\pm$  8&22.29$\pm$.01&22.61$\pm$.02&23.27$\pm$.03&23.4$\pm$.1&Yes\\
p2\_HII\_20      &12:43:40.29&16:24:01.1& 864$\pm$  3&22.54$\pm$.11&22.74$\pm$.12&22.73$\pm$.14&23.9$\pm$.1&Yes\\
p2\_HII\_57      &12:43:36.40&16:23:26.8& 959$\pm$  5&22.88$\pm$.11&22.95$\pm$.07&23.97$\pm$.22&23.8$\pm$.1&Yes\\
p2\_HII\_26      &12:43:43.10&16:23:00.7& 795$\pm$ 11&22.16$\pm$.04&22.67$\pm$.09&23.98$\pm$.44&23.8$\pm$.1&Yes\\
p2\_HII\_29      &12:43:50.54&16:24:49.4& 664$\pm$ 11& \textsuperscript{***}& \textsuperscript{***}& \textsuperscript{***}&24.7$\pm$.1&Yes\\
p2\_gal\_172     &12:43:34.89&16:23:10.1& 975$\pm$  7&24.38$\pm$.08&24.07$\pm$.05&24.94$\pm$.14&25.8$\pm$.1&Yes\\
p2\_gal\_180     &12:43:48.79&16:23:09.0& 659$\pm$ 12& ---& ---& ---&26.8$\pm$.1&Yes\\
p2\_gal\_200     &12:43:43.78&16:22:37.9& 808$\pm$ 13&24.93$\pm$.19&25.68$\pm$.37&26.51$\pm$.91&25.4$\pm$.1&Yes\\
\hline
&&&&GCs&&&&\\
\hline
p1\_HST\_0       &12:43:36.78&16:23:28.8& 641$\pm$ 20&21.51$\pm$.01&21.00$\pm$.01&20.69$\pm$.01&24.0$\pm$.1&No \\
p1\_sup\_5       &12:43:44.17&16:24:19.6& 912$\pm$ 51&22.70$\pm$.06&21.84$\pm$.02&21.50$\pm$.02&25.4$\pm$.1&No \\
p1\_sup\_6       &12:43:46.98&16:23:12.1& 738$\pm$ 31&22.36$\pm$.07&21.88$\pm$.03&21.51$\pm$.02&24.9$\pm$.1&Yes\\
p1\_sup\_10      &12:43:43.46&16:24:00.4& 895$\pm$ 39&21.86$\pm$.11&21.75$\pm$.12&21.38$\pm$.09&25.4$\pm$.1&No \\
p1\_sup\_27      &12:43:34.10&16:22:59.6&1028$\pm$ 40&22.00$\pm$.01&21.51$\pm$.01&21.24$\pm$.01&24.6$\pm$.1&Yes\\
p1\_sup\_39      &12:44:01.51&16:24:10.1& 821$\pm$ 38&21.92$\pm$.01&21.39$\pm$.01&21.14$\pm$.01&24.4$\pm$.1&No \\
p1\_sup\_63      &12:43:22.76&16:24:05.9& 861$\pm$ 49&22.73$\pm$.01&22.19$\pm$.01&21.96$\pm$.01&25.2$\pm$.1&Yes\\
p1\_sup\_88      &12:43:44.76&16:22:23.4& 801$\pm$ 40&23.25$\pm$.02&22.55$\pm$.01&22.14$\pm$.01&25.7$\pm$.1&Yes\\
p1\_sup\_121     &12:43:47.85&16:24:02.2& 659$\pm$ 67&23.29$\pm$.04&22.80$\pm$.02&22.39$\pm$.02&25.8$\pm$.1&Yes\\
p1\_sup\_188     &12:43:51.66&16:23:35.4& 641$\pm$ 12&20.95$\pm$.01&20.41$\pm$.01&20.14$\pm$.01&23.5$\pm$.1&Yes\\
p2\_sup\_123     &12:43:34.49&16:23:35.7& 986$\pm$ 63&23.39$\pm$.03&22.55$\pm$.01&22.15$\pm$.01&25.8$\pm$.1&Yes\\
p2\_sup\_170     &12:43:27.46&16:24:08.8&1011$\pm$ 41&22.98$\pm$.01&22.56$\pm$.01&22.31$\pm$.01&25.5$\pm$.1&Yes\\
p2\_sup\_283     &12:43:25.85&16:23:50.6& 974$\pm$ 14&20.92$\pm$.01&20.51$\pm$.01&20.29$\pm$.01&23.4$\pm$.1&Yes\\
\hline\\
\multicolumn{4}{l}{\textsuperscript{*}\footnotesize{Measured from the higher resolution long slit spectrum.}}\\
\multicolumn{4}{l}{\textsuperscript{**}\footnotesize{Identified by eye on image.}}\\
\multicolumn{4}{l}{\textsuperscript{***}\footnotesize{Photometry affected by crowding.}}\\
\end{tabular}
\end{center}
\end{table*}

\section*{Acknowledgments}
We thank M. Fardal and J. C. Mihos for kindly providing their $N$-body simulations, and Raphael Sadoun and Robyn Sanderson for helpful comments. CF acknowledges co-funding under the Marie Curie Actions of the European Commission (FP7-COFUND). HL acknowledges a fellowship from the European Commission's Framework Programme 7, through the Marie Curie Initial Training Network CosmoComp (PITN-GA-2009-238356). This work was supported by National Science Foundation grants AST-0909237 and AST-1211995. Part of the data presented herein were obtained at the W.M. Keck Observatory, which is operated as a scientific partnership among the California Institute of Technology, the University of California and the National Aeronautics and Space Administration. The Observatory was made possible by the generous financial support of the W.M. Keck Foundation. The analysis pipeline used to reduce the DEIMOS data was developed at UC Berkeley with support from NSF grant AST-0071048. Based in part on data collected at Subaru Telescope, which is operated by the National Astronomical Observatory of Japan.  Based in part on data collected at Subaru Telescope (which is operated by the National Astronomical Observatory of Japan), via Gemini Observatory time exchange (GN-2010B-C-204). The {\it GALEX} NUV image presented in this paper was obtained from the Mikulski Archive for Space Telescopes (MAST). STScI is operated by the Association of Universities for Research in Astronomy, Inc., under NASA contract NAS5-26555. Support for MAST for non-{\it HST} data is provided by the NASA Office of Space Science via grant NNX13AC07G and by other grants and contracts.
\bibliographystyle{mn2e}
\bibliography{biblio}

\label{lastpage}

\end{document}